\documentclass[aps,pra,floats, groupedaddress, showpacs, superscriptaddress, twocolumn, 10pt]{revtex4}%
\usepackage{amsmath}
\usepackage{amsfonts}
\usepackage{amssymb}
\usepackage{graphicx, subfigure}
\usepackage{dcolumn}
\usepackage{natbib}
\usepackage{color}

\begin{document}

\title{Non-Gaussian entangled states and quantum teleportation of Schr{\"o}dinger-cat states\footnote{A review in honor of V. Manko and M. Manko}}
\author{Kaushik P.\ Seshadreesan}
\email{ksesha1@lsu.edu}
\affiliation{Hearne Institute for Theoretical Physics and Department of Physics and Astronomy, Louisiana State University, Baton Rouge, LA 70803, USA}

\author{Jonathan P.\ Dowling}
\affiliation{Hearne Institute for Theoretical Physics and Department of Physics and Astronomy, Louisiana State University, Baton Rouge, LA 70803, USA}
\affiliation{Beijing Computational Science Research Center, Beijing, 100084, China}

\author{Girish S. Agarwal}
\affiliation{Department of Physics, Oklahoma State University, Stillwater, OK 74078, USA}

\date{\today}

\begin{abstract}

In continuous-variable quantum information, non-Gaussian entangled states that are obtained from Gaussian entangled states via photon subtraction are known to contain more entanglement. This makes them better resources for quantum information processing protocols, such as, quantum teleportation. We discuss the teleportation of non-Gaussian, non-classical Schr{\"o}dinger-cat states of light using two-mode squeezed vacuum light that is made non-Gaussian via subtraction of a photon from each of the two modes. We consider the experimentally realizable cat states produced by subtracting a photon from the single-mode squeezed vacuum state. We discuss two figures of merit for the teleportation process, a)~the fidelity, and b)~the maximum negativity of the Wigner function at the output. We elucidate how the non-Gaussian entangled resource lowers the requirements on the amount of squeezing necessary to achieve any given fidelity of teleportation, or to achieve negative values of the Wigner function at the output.
\end{abstract}

\pacs{270.0270, 270.6570, 270.5585}

\maketitle

\section{Introduction}

Entanglement is a vital resource for quantum information processing. In continuous-variable quantum information, a commonly used form of entanglement is the two-mode squeezed vacuum state produced by the process of down conversion. A strong pump emitting photons at frequency $\omega_p$ interacts with a nonlinear crystal containing a second order nonlinearity, generating pairs of photons at frequencies $\omega_a$ and $\omega_b$, such that $\omega_a+\omega_b=\omega_p$. The quantum state at the output of the down-conversion process is given by
\begin{align}
\label{tmsv}
|\xi\rangle&=\hat{S}(\xi)|0\rangle_a|0\rangle_b \ (\xi=r e^{i \phi})\nonumber\\
&=\frac{1}{\cosh r}\sum_{0}^{\infty}e^{in\phi}(\tanh r)^n|n\rangle_a|n\rangle_b,
\end{align}
where $\hat{S}(\xi)=\exp{\left(\xi \hat{a}^{\dagger}\hat{b}^{\dagger}-\xi^{*}\hat{a}\hat{b}\right)}$ is the two-mode squeezing operator~\cite{gerryknight}, $r$ and $\phi$ are the squeezing amplitude and phase, respectively, and $\hat{a}$ and $\hat{b}$ are the mode operators of the two-mode output field. An important property of this state is that its Wigner function is Gaussian. The squeezing associated with this state is $(1-\exp(-2r))/2$, or $-10\log_{10}\exp(-2r)$ dB. The entanglement in the state, measured in terms of the logarithmic negativity, is given by  $\varepsilon_{\xi}={\rm log}_2 \left(e^{2r}\right)$. Thus, both entanglement as well as the squeezing are determined by the parameter $r$, which is proportional to the amplitude of the pump field and the second order nonlinearity of the crystal. It is not easy to obtain large values of $r$ unless one does down conversion in a resonant cavity. In the limit of $r$ becoming infinitely large, the two-mode squeezed vacuum state tends towards the ideal EPR state~\cite{EPR_35}. Despite leaps in technological advancement, the state-of-the-art vacuum squeezing for the two-mode squeezed vacuum state, however, remains about $10$ dB~\cite{Eberle_13, Vahlbruch_08}. Hence, techniques that improve the performance of quantum information processing without demanding higher magnitudes of squeezing are vital commodities. In this regard, for any given amount of squeezing, the two-mode squeezed vacuum state which is made non-Gaussian via photon addition or subtraction~\cite{Agarwal_91}, is known to contain more entanglement than the two-mode squeezed vacuum state~\cite{Kitagawa_06, Agarwal_11, Agarwal_12}. Hence, non-Gaussian entangled states should be better suited for quantum information protocols~\cite{opatrny_00, Cochrane_02, Olivares_03, Olivares_04, DAnno_07}. 

In this review, we introduce the non-Gaussian entangled states obtained by subtracting one photon from each mode of the two-mode squeezed vacuum state. We present a scheme that generates such states, and discuss the important properties and distinct features of the states. We discuss the advantage of such states in continuous-variable quantum information protocols. We focus on the teleportation protocol~\cite{Bennett_93,Vaidman_94, Braunstein_98}, and discuss the teleportation of non-Gaussian quantum states such as the cat states, which are states of the form
\begin{equation}
\label{catex}
|\Psi_{\rm cat}\rangle=N^{-1}\left(|+\alpha_0\rangle+e^{i \theta}|-\alpha_0\rangle\right),
\end{equation}
where $N=\sqrt{2\left(1+e^{-2|\alpha_0|^2}\cos\theta\right)}$ and $|\pm\alpha_0\rangle$ are coherent states. States of the form given in Eq.~(\ref{catex}) with $\theta =0$ and $\theta=\pi$ were introduced by Dodonov, Malkin and Manko in ref.~\cite{DMM74} under the names of ``even and odd coherent states".

The teleportation of Gaussian states, such as the coherent state, has been demonstrated in many experiments since 1998 with high fidelities~\cite{fidnote, Furasawa_98, Bowen_03, Yonezawa_04}. However, the first-ever teleportation of a non-Gaussian state was carried out rather recently~\cite{Lee_11}. Using the standard protocol for continuous variable teleportation due to Vaidman, Braunstein and Kimble~\cite{Vaidman_94, Braunstein_98}, with the two-mode squeezed vacuum state as the entangled resource, Lee {\it et al.} teleported a cat-like state of the form 
\begin{equation}
\label{catlikex}
|\Phi_{\rm cat}\rangle=\frac{1}{\sinh r} \hat{a}\hat{S}(\xi)|0\rangle, \ \xi=\rho e^{i \varphi},
\end{equation}
where $\hat{S}(\xi)=\exp{\left(\left(\xi \hat{a}^{\dagger 2}-\xi^{*}\hat{a}^2\right)/2\right)}$~\cite{gerryknight} is the single-mode squeezing operator acting on the mode $\hat{a}$. An input cat-like state $|\Phi_{\rm cat}\rangle$ of Eq.~(\ref{catlikex}) of $75\pm 0.5\%$ fidelity with respect to the cat state $|\Psi_{\rm cat}\rangle$ of coherent amplitude $|\alpha|^2\approx 1$ and $\theta=\pi$ was teleported, achieving an output fidelity of $45\pm 1\%$ with respect to the same cat state~\cite{iofidnote, Lee_11, Grangier_11}. Lee {\it et al.} also observed that, the negativity of the input Wigner function remained preserved at the output in the experiment, which confirmed the transfer of the non-Gaussianity of the state from the input to the output. Further, they verified that the maximum negativity of the output Wigner function was in good agreement with the prediction based on a model for non-unity gain teleportation given by Mista {\it et al.}~\cite{Mista_10}.

The non-Gaussian entangled states obtained from photon subtraction also find other applications in continuous-variable quantum information apart from quantum teleportation.~These include loophole free tests for Bell inequality violations using homodyne detection~\cite{NC04, GFCW04, GPFC05, DK05}, quantum bit commitment that is robust against Gaussian attacks~\cite{MMLC10}, and quantum optical interferometry for sub-shot-noise phase estimation (super sensitivity) and sub-Rayleigh spatial resolution (super resolution) with potential use in sensing, imaging and lithography~\cite{CG12}.

The non-Gaussian entangled states discussed here are not the only interesting instances of such states. Another well-known example of an entangled state that is non-Gaussian~\cite{Dodo02} is the $N00N$ state~\cite{Dowling08} $(|N\rangle|0\rangle + |0\rangle|N\rangle)/\sqrt{2}$, which has many desirable quantum properties. In optical interferometry, the $N00N$ state is capable of super sensitivity and super resolution. More generally one could think of a combination of two different Fock states with $N$ and $M$ photons, and construct a state like $(|N\rangle|M\rangle + |M\rangle|N\rangle)/\sqrt{2}$~\cite{HWD08}. Instances of such states have been shown to offer similar benefits as the $N00N$ state in optical interferometry, while being more robust against photon loss than the latter~\cite{JBWKLD12, RJD13}. However, the production of such states remains hard. Another important class of entangled non-Gaussian states are the so-called vortex states~\cite{AB06, SA00, APR97}, which can be obtained from $|N\rangle|M\rangle$ by rotations of the form
\begin{equation}
\exp(i\pi(\hat{a}^\dagger\hat{b}+\hat{b}^\dagger\hat{a})/4).
\end{equation}
These states can be produced by launching single photons on beam splitters, or in two coupled waveguides. A good way to test their quantum character is via entropic uncertainty relations~\cite{CPMA14}.

\section{Non-Gaussian entanglement}

In this section, we discuss the generation, properties and entanglement content of the two-mode squeezed vacuum state that is made non-Gaussian via photon subtraction. But first of all, let us review some properties of the Gaussian two-mode squeezed vacuum state.

\subsection{The two-mode squeezed vacuum state}
The two-mode squeezed-vacuum state $|\xi\rangle$ is defined as given in Eq.~(\ref{tmsv}). The two-mode squeezing operation is implemented, e.g., in a non-degenerate parametric downconversion process, which involves pumping a $\chi^{(2)}$ nonlinear optical device with photons of frequency $\omega_p$, some of which then get converted into a pair of photons---of frequencies $\omega_a$ and $\omega_b$, respectively, such that $\omega_p=\omega_a+\omega_b$.

The Wigner function of the two-mode squeezed vacuum state is Gaussian, and is given by
\begin{align}
W_{\xi}(\alpha,\beta)&=\frac{4}{\pi^2}\exp\left(-2|\alpha\cosh r-\beta^*\sinh r\ e^{i\phi}|^2\right)\nonumber\\
&\times\exp\left(-2|-\alpha^*\sinh r\ e^{i\phi}-\beta\cosh r|^2\right).
\end{align}
For quadrature operators  $\hat{x}_a$, $\hat{y}_a$, $\hat{x}_b$, and $\hat{y}_b$ defined in terms of the mode operators $\hat{a}$, $\hat{a}^\dagger$, $\hat{b}$, and $\hat{b}^\dagger$ as
\begin{eqnarray}
\label{homodyne2}
\hat{x}_a=\frac{1}{\sqrt{2}}\left(\hat{a}+\hat{a}^{\dagger}\right),\ \ \hat{y}_a=\frac{1}{\sqrt{2}i}\left(\hat{a}-\hat{a}^{\dagger}\right),\nonumber\\
\hat{x}_b=\frac{1}{\sqrt{2}}\left(\hat{b}+\hat{b}^{\dagger}\right),\ \ \hat{y}_b=\frac{1}{\sqrt{2}i}\left(\hat{b}-\hat{b}^{\dagger}\right),
\end{eqnarray}
the quadrature distribution of the two-mode squeezed vacuum state is given by~\cite{Agarwal_12}
\begin{align}
\label{quadtmsv}
\psi_{\xi}(x_a,x_b)&=\frac{1}{\sqrt{(1-\eta^2)\pi\cosh^2 r}}\exp\left(-\frac{1}{2}(x_a^2+x_b^2)\right)\nonumber\\
&\exp\left(\frac{2x_a x_b \eta -(x_a^2+x_b^2)\eta^2}{1-\eta^2}\right),
\end{align}
where $x_a$, $x_b$ are the eigenvalues values of the corresponding quadrature operators, and $\eta=e^{i\phi} \tanh r$.

The squeezing associated with the two-mode squeezed vacuum state is given by
\begin{equation}
S_{\frac{\phi}{2}+\frac{\pi}{2}}=-1/2(1-e^{-2r})\leq 0,
\end{equation}
where it is calculated as
\begin{align}
\label{sqpara}
S_\theta&=\left(\Delta X_\theta^{(d)}\right)^2-1/2,\nonumber\\
X_\theta^{(d)}&=(\hat{d}e^{-i\theta}+\hat{d}^\dagger e^{i\theta})/\sqrt{2},\ \ \hat{d}=(\hat{a}+\hat{b})/\sqrt{2}.
\end{align}

The entanglement content of the two-mode squeezed vacuum state can be characterized using the logarithmic negativity measure~\cite{Vidal_02, Plenio_05}, which is known to be a good measure of entanglement for CV states. It is defined as
\begin{equation}
\label{lognegdef}
\varepsilon={\rm log}_2\left(1+2\mathcal{N}\right),
\end{equation}
where $\mathcal{N}$ is the absolute value of the sum of all negative eigenvalues associated with the partial transpose of the density operator. The logarithmic negativity of the two-mode squeezed vacuum state is found to be~\cite{Agarwal_11}
\begin{equation}
\label{ephtmsvlogneg}
\varepsilon_{\xi}={\rm log}_2 \left(e^{2r}\right).
\end{equation}


\subsection{The non-Gaussian two-photon-subtracted two-mode squeezed vacuum state}
\begin{figure}[h]\centering
\includegraphics[scale=0.75]{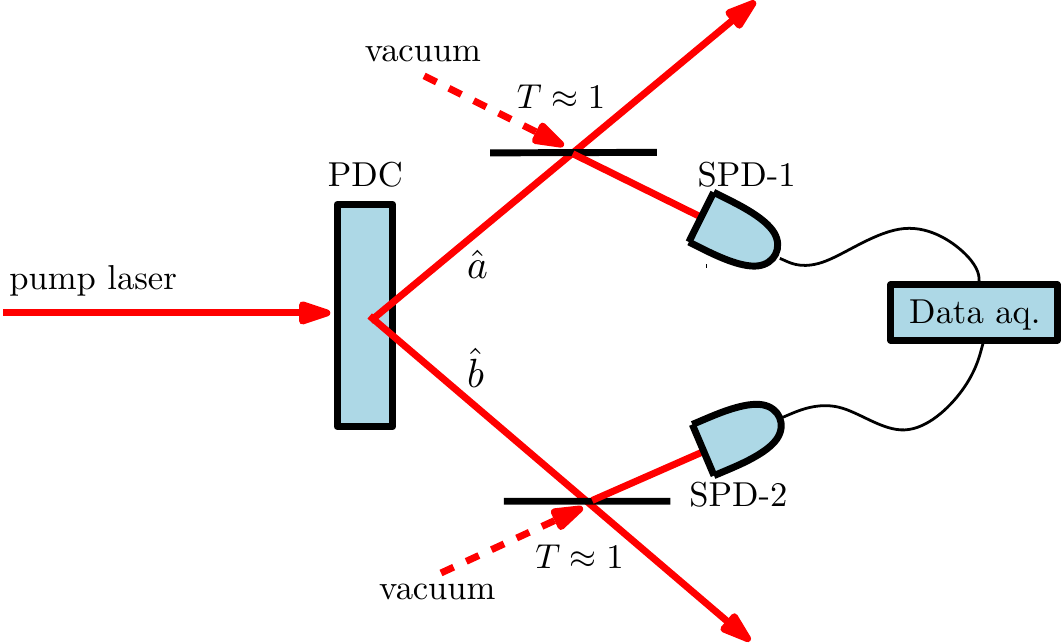}
\caption{(Color online) A conditional measurement scheme, wherein a photon is subtracted from each of the two modes of the two-mode squeezed vacuum state. PDC stands for parametric down-conversion, and the SPDs are single photon detectors.}
\label{schematic}
\end{figure}

The two-photon-subtracted two-mode squeezed vacuum state is defined as
\begin{eqnarray}
&|\xi\rangle_{{\rm TPS}}\propto \hat{a}\hat{b}\hat{S}(\xi)|0\rangle|0\rangle\\
&=\frac{1}{\sqrt{1+\tanh^2 r}}\hat{S}\left(\xi\right)\left(|0\rangle_a|0\rangle_b+e^{i\phi}\tanh r |1\rangle_a|\ 1\rangle_b\right)\label{notoneform}\\
&=\frac{1}{\cosh^{3} r\sqrt{1+\tanh^2 r}}\sum_{n=0}^{\infty}e^{i n \phi}(\tanh r)^n (n+1) |n\rangle_a|n\rangle_b,\nonumber\\
\label{ephtmsv1}
\end{eqnarray}
where the state has been suitably normalized. This state is generated from the two-mode squeezed vacuum state given in Eq.~(\ref{tmsv}) as follows. Consider the scheme shown in Fig.~\ref{schematic}. Parametric downconversion generates the two-mode squeezed-vacuum state $|\xi\rangle$. Highly transmissive beam splitters, are placed, one in each of the two modes of the two-mode squeezed vacuum state. These beam splitters feed single-photon detectors (SPD-1, 2). When the SPDs register a coincidence detection, the two-photon-subtracted two-mode squeezed vacuum state is heralded~\cite{PZKB07, KBKL13}.

\begin{figure}\centering
\includegraphics[scale=0.6]{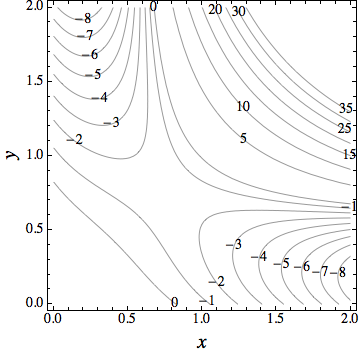}
\caption{A plot of the polynomial pre-factor in the Wigner function of the two-photon-subtracted two-mode squeezed vacuum state given in Eq.~(\ref{wignegtps}), for value of the squeezing parameter $r=1$. The labels $x$ and $y$ denote $|\tilde{\alpha}|$ and $|\tilde{\beta}|$, respectively, and we set $\tilde{\alpha}\tilde{\beta}\cos\phi=-xy$. The plot takes on negative values, which indicates that the Wigner function is negative in certain parts of phase space.}
\label{tps_neg}
\end{figure}
The Wigner function of the two-photon-subtracted two-mode squeezed vacuum state is non-Gaussian, and is given by
\begin{align}
&W_{TPS}(\alpha,\beta)=\frac{4}{(1+\tanh^2r)\pi^2}\exp\left(-2\left(|\tilde{\alpha}|^2+|\tilde{\beta}|^2\right)\right)\times\nonumber\\
&\left(1+8\tilde{\alpha}\tilde{\beta}\cos\phi\tanh r+(4|\tilde{\alpha}|^2-1)(4|\tilde{\beta}|^2-1)\tanh^2r\right),
\label{wignegtps}
\end{align}
where
\begin{equation}
\left[
\begin{array}{c}
\tilde{\alpha}\\
\tilde{\beta}^*
\end{array}
\right]= \left[
\begin{array}{cc}
\cosh r   & -\sinh r e^{i\phi} \\
-\sinh r e^{-i\phi} & \cosh r
\end{array}
\right]
\left[
\begin{array}{c}
\alpha\\
\beta^*
\end{array}
\right].
\label{TRANS}
\end{equation}
(See Appendix A for details of the calculation.) Figure~\ref{tps_neg} demonstrates the fact that this non-Gaussian Wigner function takes on negative values in certain parts of phase space. The quadrature distribution of the two-photon-subtracted two-mode squeezed vacuum is given by
\begin{align}
\label{quadtps}
&\psi_{\rm TPS}(x_a,x_b)=N^{-1}\eta(1+\eta\frac{\partial}{\partial\eta})\psi_{\xi}(x_a,x_b), \nonumber\\
 &\eta=e^{i\phi} \tanh r,\ \ N=\sinh r\cosh r\sqrt{1+\tanh^2 r},
\end{align}
where $\psi_{\xi}(x_a,x_b)$ is the quadrature distribution of the two-mode squeezed vacuum state, given in Eq.~(\ref{quadtmsv}).

\begin{figure}\centering
\includegraphics[scale=0.45]{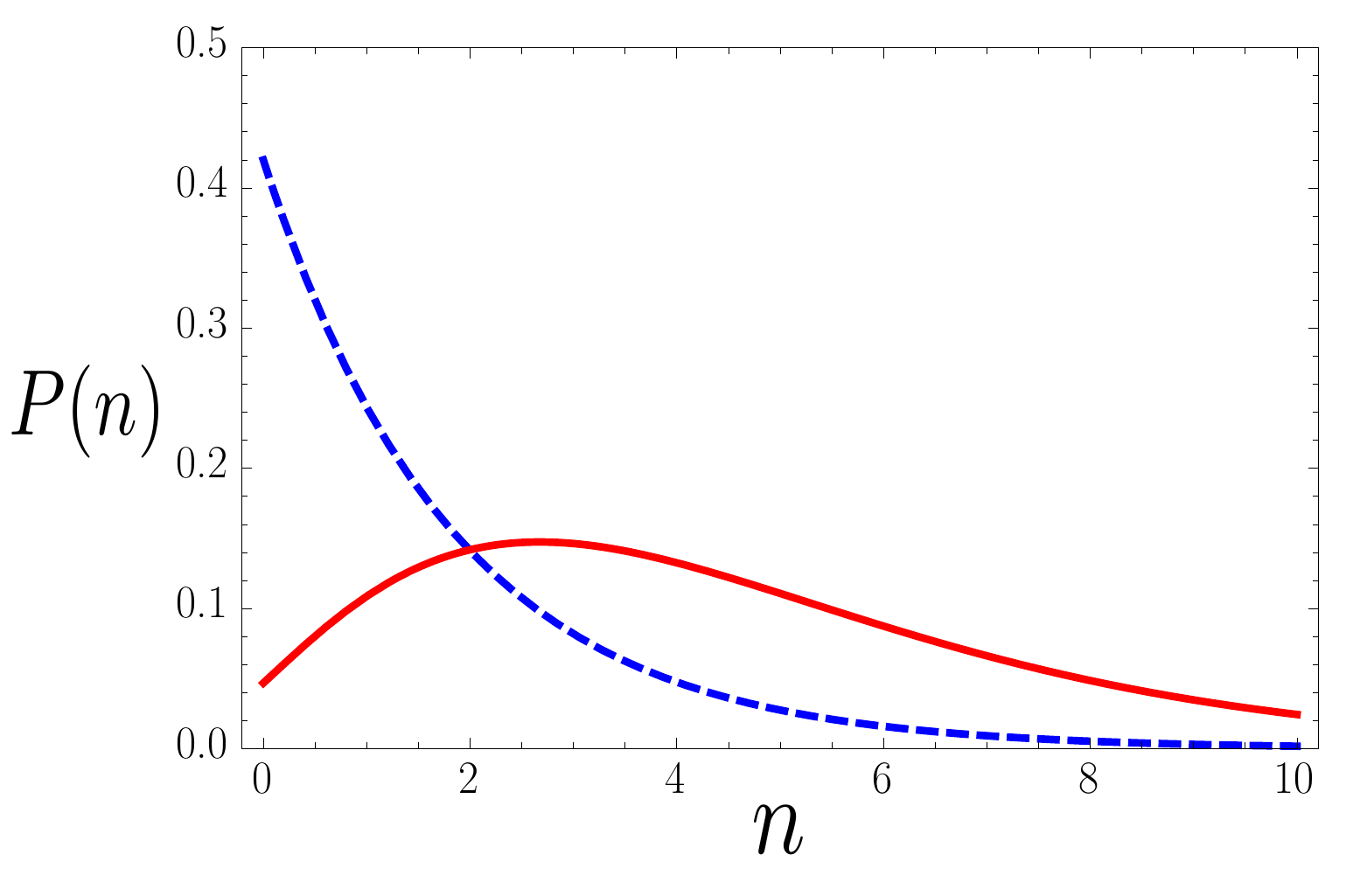}
\caption{(Color online) The photon number distributions for the two-mode squeezed vacuum state (dashed, blue) and two-photon subtracted the two-mode squeezed vacuum state (solid, red) for the squeezing parameter $r=1$. $P(n)$ corresponds to the probability of finding $n$ photons in each of the two modes simultaneously.}
\label{pndisr1}      
 \end{figure}

  \begin{figure*}\centering
        \subfigure[]{%
           \label{tquad}
           \includegraphics[scale=0.32]{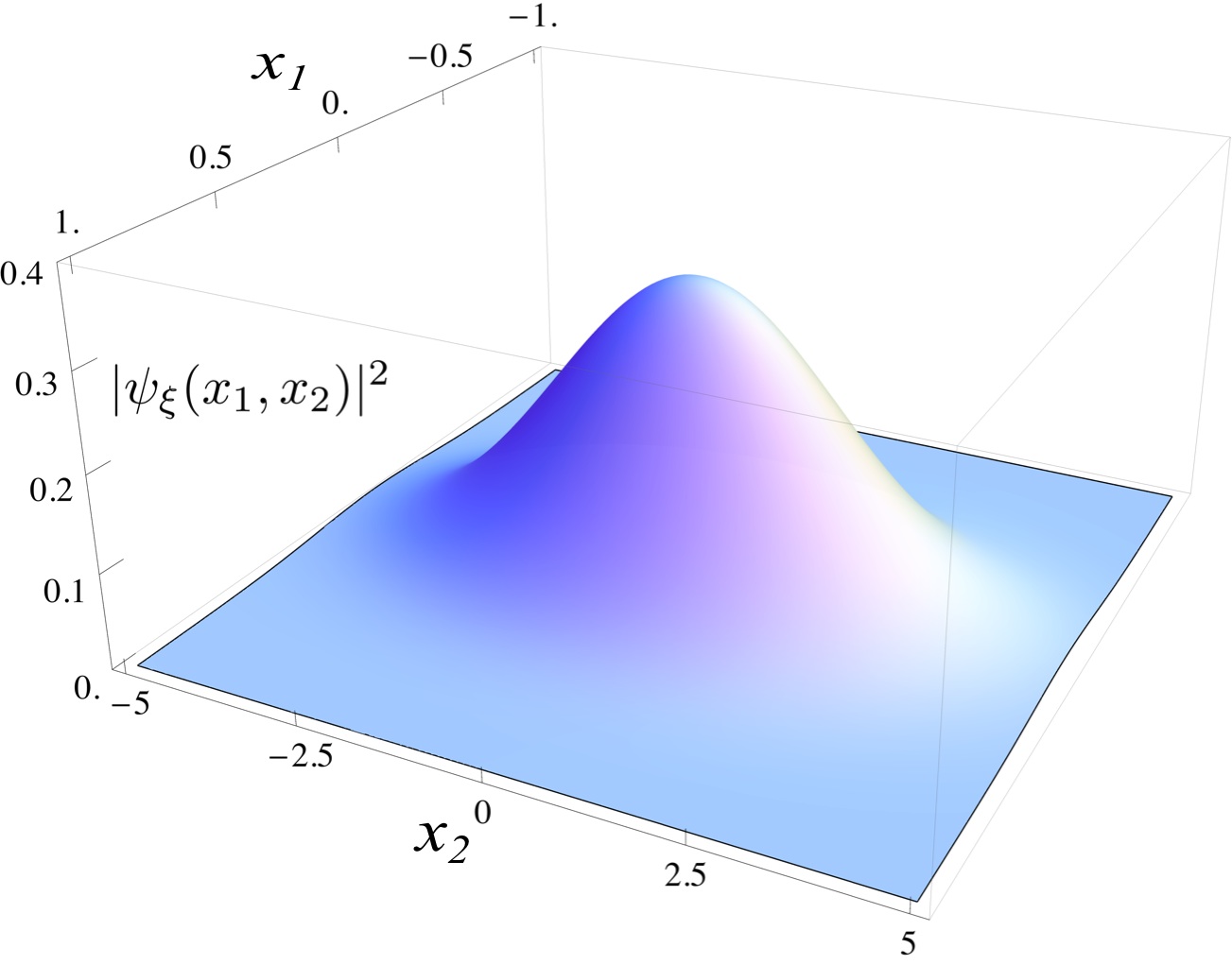}
        }
           \subfigure[]{%
           \label{tquadr5}
           \includegraphics[scale=0.32]{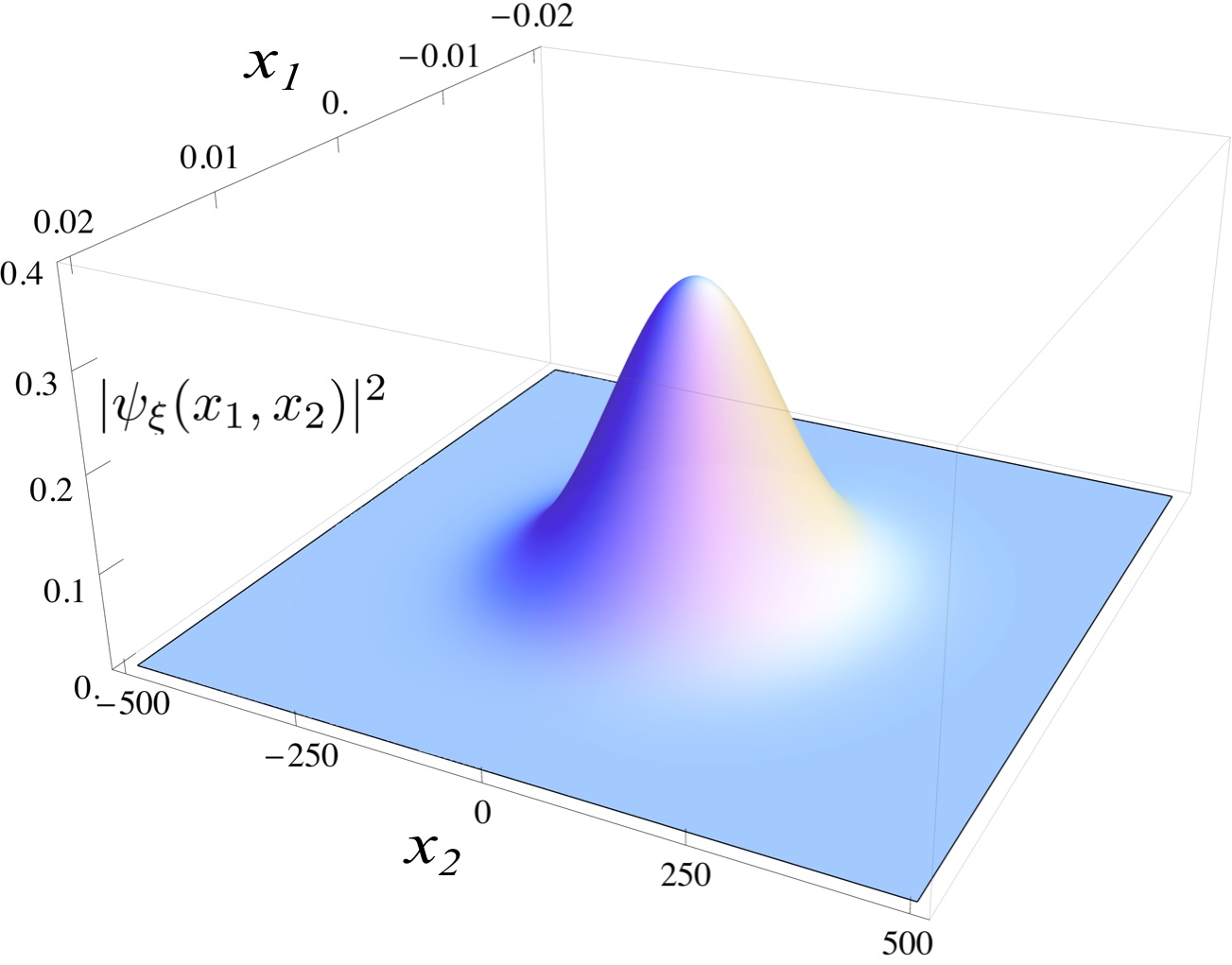}
        }
        \\
        \subfigure[]{%
           \label{equadr1}
           \includegraphics[scale=0.32]{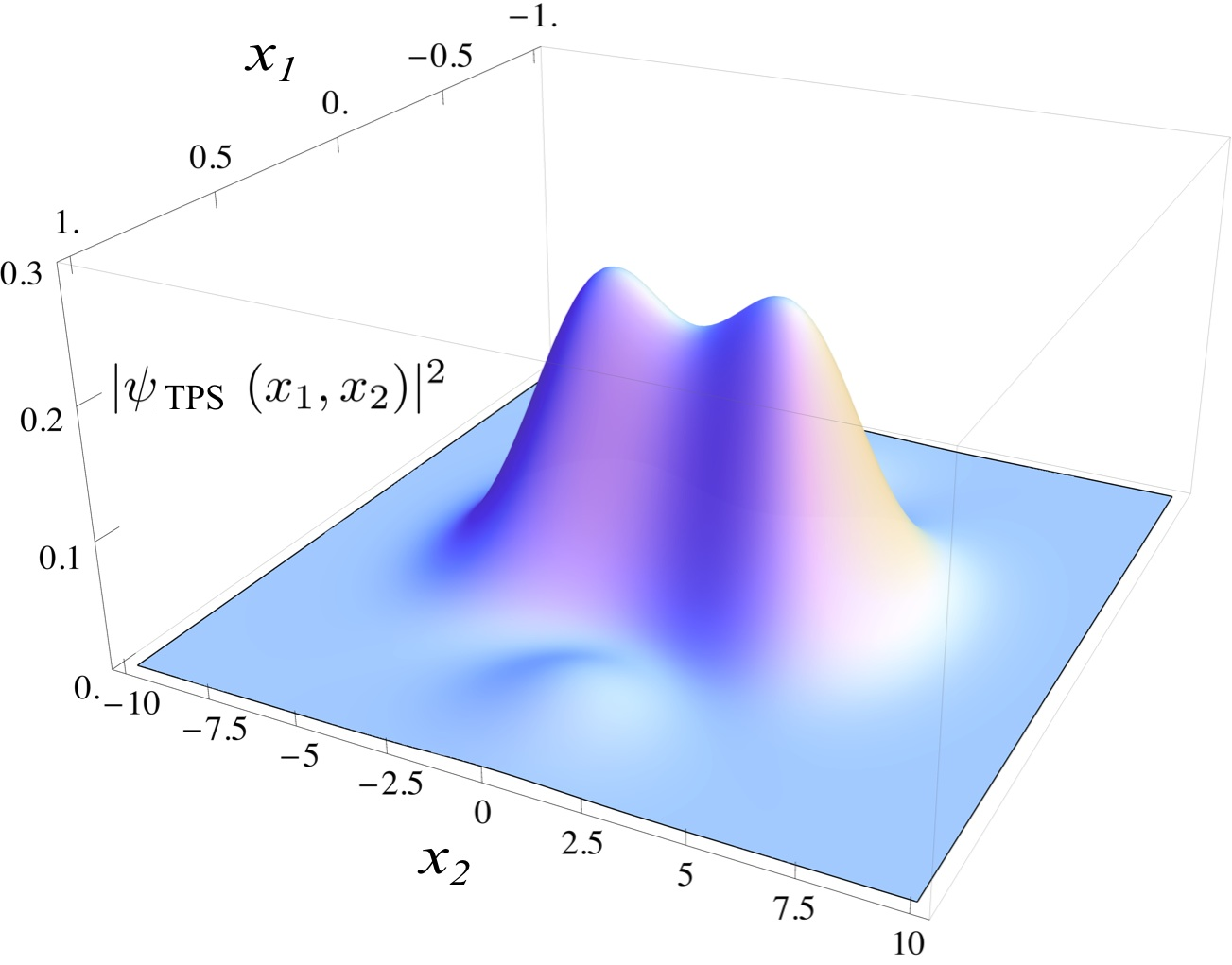}
        } 
        \subfigure[]{%
            \label{equadr5}
            \includegraphics[scale=0.32]{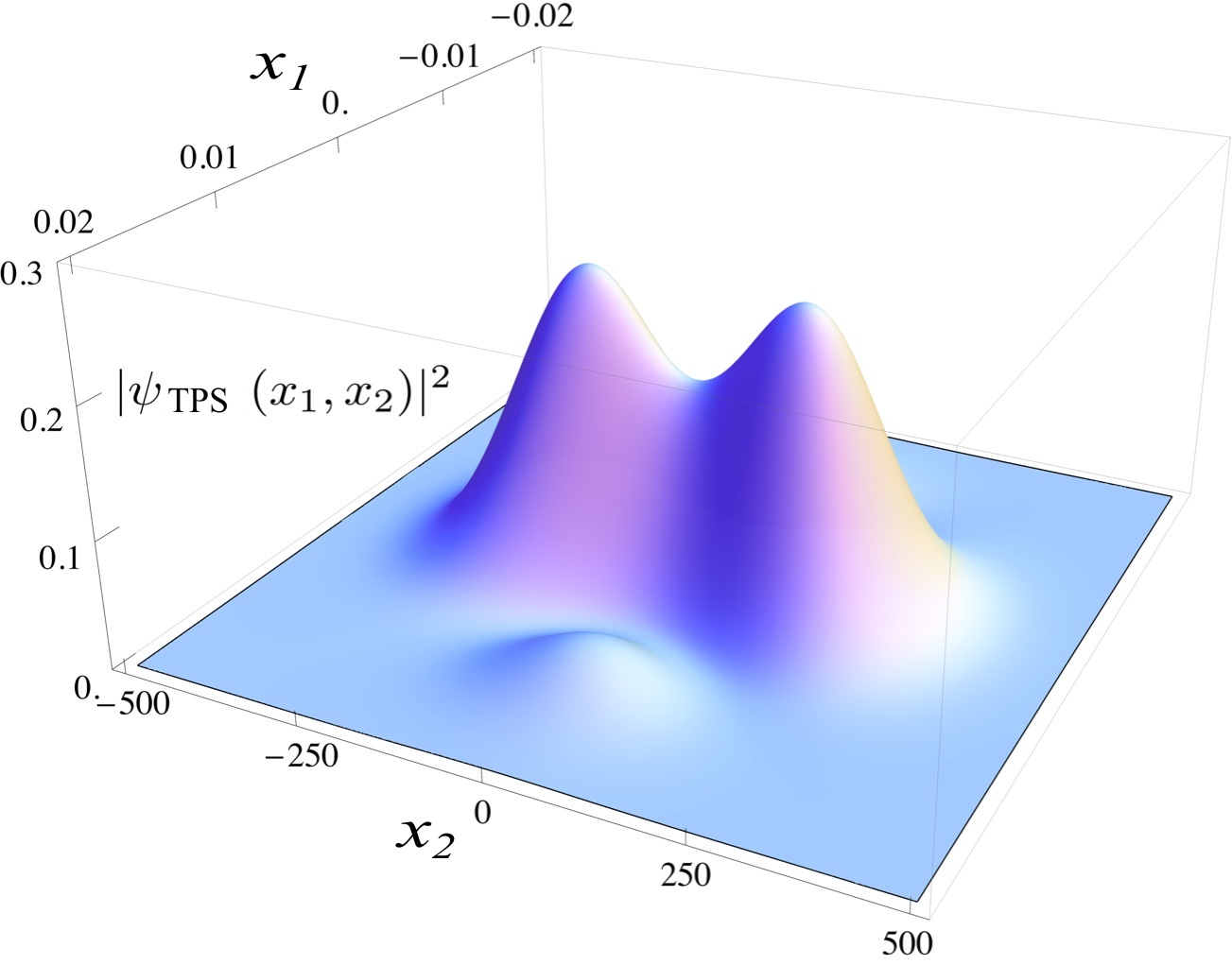}
        }%
\caption{(Color online) (a) and (b) The intensity $|\psi_{\xi}(x_1,x_2)|^2$ as a function of $x_1$ and $x_2$, when the squeezing parameters are chosen to be $r=1,\ \phi=\pi$, and $r=5,\ \phi=\pi$, respectively. (c) and (d) The intensity $|\psi_{\rm TPS}(x_1,x_2)|^2$ as a function of $x_1$ and $x_2$ when the squeezing parameters are chosen to be $r=1,\ \phi=\pi$, and $r=5,\ \phi=\pi$, respectively.}
\end{figure*}

 \begin{figure}\centering
\includegraphics[scale=0.55]{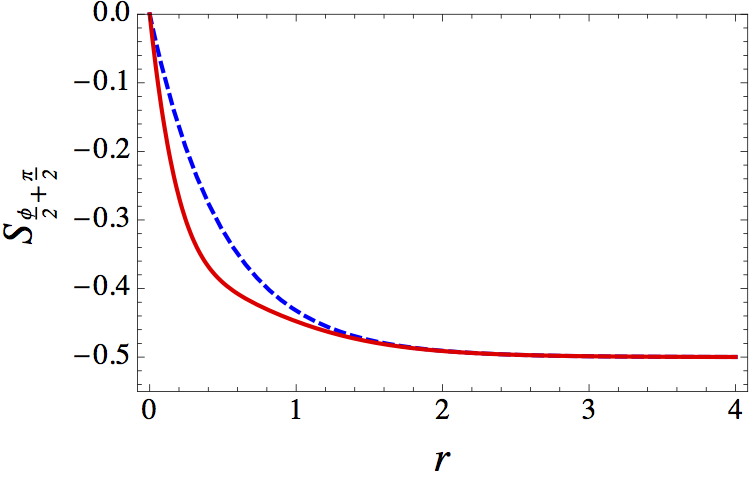}
\caption{(Color online) The squeezing parameter of the two-mode squeezed vacuum state (dashed, blue) and two-photon subtracted the two-mode squeezed vacuum state (solid, red).}
\label{sqparafig}      
 \end{figure}

\begin{figure*}\centering
\includegraphics[scale=0.75]{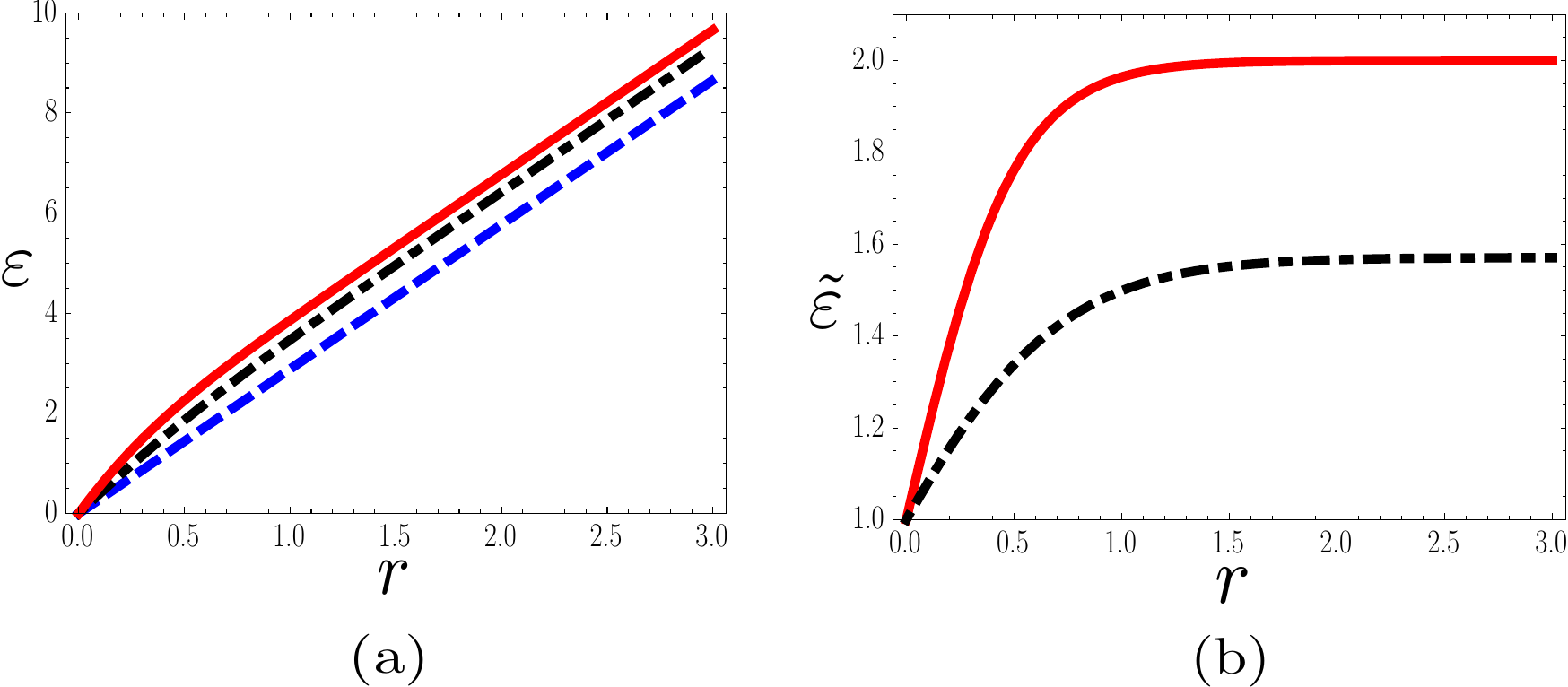}
\caption{(Color online) (a) The log-negativity parameter $\varepsilon$ for the two-mode squeezed vacuum state (dashed, blue), photon-added (dot-dashed, black), and two-photon-subtracted (solid, red) the two-mode squeezed vacuum state as a function of the squeezing parameter $r$. (b) The ratio $\tilde{\boldsymbol{\varepsilon}}$ for photon-added (dot-dashed, black) and two-photon-subtracted (solid, red) the two-mode squeezed vacuum state as a function of the squeezing parameter $r$.}
\label{logneg}
\end{figure*}

Squeezing in the two-photon-subtracted two-mode squeezed vacuum state, calculated similarly to Eq.~(\ref{sqpara}), is found to be:
\begin{align}
S_{\pi}=\frac{1}{2} e^{-r} \left(e^r-e^{-r}\right) \left(\frac{e^{2 r}-e^{-2 r}}{e^{-2 r}+e^{2 r}}-2\right).
\end{align}
Further the entanglement content of the two-photon-subtracted two-mode squeezed vacuum state based on the logarithmic negativity defined in Eq.~(\ref{lognegdef}) is found to be $\varepsilon_{\rm TPS}={\rm log}_2 \left(e^{4r}/\cosh 2r\right)$. (See Appendix B for the derivation.)

It is interesting to note that the two-photon-subtracted two-mode squeezed vacuum state, upon a basis change, can be shown to be a cat-like state. Consider the following new basis:
\begin{align}
\label{basischange}
\hat{a}_+=\frac{1}{\sqrt{2}}\left(\hat{a}+\hat{b}\right),\ \ \hat{a}_-=\frac{1}{\sqrt{2}}\left(\hat{a}-\hat{b}\right).
\end{align}
In this new basis, the two-mode squeezing operator becomes
\begin{equation}
\hat{S}(\xi)=\hat{S'}_+(\xi)\hat{S'}_-(-\xi),
\end{equation}
where $\hat{S'}_+(\xi)=\exp{\left(\xi \hat{a}^{\dagger 2}_+-\xi^{*}\hat{a}^2_+\right)}$ is the single-mode squeezing operator, and likewise $\hat{S'}_-(\xi)$ for the other mode. Further, the two-photon-subtracted two-mode squeezed vacuum state becomes:
\begin{align}
\left(\hat{a}_+^2\hat{S'}_+(\xi)|0\rangle\right)\left(\hat{S'}_-(\xi)|0\rangle\right)-\left(\hat{S'}_+(\xi)|0\rangle\right)\left(\hat{a}_-^2\hat{S'}_-(\xi)|0\rangle\right).
\end{align}

\subsection{Characteristics of the non-Gaussian two-photon-subtracted two-mode squeezed vacuum state}

We now elucidate the characteristic properties of the non-Gaussian two-photon-subtracted two-mode squeezed vacuum state in comparison to those of the two-mode squeezed vacuum state.

\subsubsection{Photon number distribution}

Figure~\ref{pndisr1} shows plots of the photon number distributions of the two states for the squeezing parameter $r=1$. We notice that the two-photon-subtracted two-mode squeezed vacuum state has a higher weighting for large photon numbers than the two-mode squeezed vacuum state---an observation made by Cochrane {\it et al.} in Ref.~\cite{Cochrane_02}. Consequently, the probability of measuring single photon coincidences in the two modes of the state $|\xi\rangle_{{\rm TPS}}$ is substantially reduced compared to that of the two-mode squeezed vacuum state. Particularly, this effect is found to grow larger as the value of the squeezing paramter $r$ is increased. The ratio $P_{\rm TPS }(1)/P_{\rm TMSV}(1)$ for the squeezing parameter $r=1$ (shown in Fig.~\ref{pndisr1}) is about $0.3$, while its value dwindles to $\approx 6\times10^{-8}$ when $r=5$.

\subsubsection{Quadrature distribution and squeezing}
 
As for the quadrature distributions, it is advantageous to work with a rotated set of coordinates $x_1$ and $x_2$, defined as
\begin{equation}
\label{coortrans}
x_1=\frac{x_a+x_b}{\sqrt{2}},\ \ x_2=\frac{x_a-x_b}{\sqrt{2}}.
\end{equation}
In terms of $x_1$ and $x_2$, the state $|\xi\rangle$ has quite a transparent structure in its quadrature distribution
\begin{eqnarray}
\label{xix1x2}
&\psi_{\xi}(x_1,x_2)=\frac{1}{\sqrt{(1-\eta^2)\pi\cosh^2 r}}e^{-\frac{1}{2}\left(\frac{1-\eta}{1+\eta}\right)x_1^2}e^{-\frac{1}{2}\left(\frac{1+\eta}{1-\eta}\right)x_2^2}.\nonumber\\
\end{eqnarray}
When $\phi=\pi$, as is well known, Eq.~(\ref{xix1x2}) has the EPR form with a narrow peak at $x_1=0$.
\begin{equation}
\label{xix1x2epr}
\psi_{\xi}(x_1,x_2)=\frac{1}{\sqrt{\pi}}e^{-x_1^2/(2e^{-2r})}e^{-x_2^2/(2e^{2r})}.
\end{equation}
Fig.~3 shows the intensities $|\psi_{\xi}(x_1,x_2)|^2$ and $|\psi_{\rm TPS}(x_1,x_2)|^2$ plotted as functions of $x_1$ and $x_2$ for values of the squeezing parameter $r=1$ and $r=5$. We find that the plot of $|\psi_{\rm TPS}(x_1,x_2)|^2$ shows a dip at $x_2=0$. Since $x_2\propto x_a-x_b$, the dip signifies a decrease in the probability of finding photons at the same values of position quadratures in the two modes for the state $|\xi\rangle_{{\rm TPS}}$. Between Figs.~\ref{equadr1} and \ref{equadr5}, we find that the dip at $x_2=0$ is even more pronounced in the latter, which corresponds to the larger value of squeezing parameter $r$.

Figure~\ref{sqparafig} shows a comparison of the squeezing $S_{\frac{\phi}{2}+\frac{\pi}{2}}$, defined in Eq.~(\ref{sqpara}), of the two-photon-subtracted two-mode squeezed vacuum state and the two-mode squeezed vacuum state. We see that the former contains more squeezing than the latter for any value of squeezing parameter $r$, until the values corresponding to both converge to $-1/2$.

\subsubsection{Entanglement in terms of logarithmic negativity}

Figure~\ref{logneg} (a) shows a plot of the logarithmic negativities of  the two-photon-subtracted two-mode squeezed vacuum state, the single-photon-added two-mode squeezed vacuum state~\cite{spatmsv} and the two-mode squeezed vacuum state, plotted as a function of the squeezing parameter $r$. As one can see, in terms of log-negativity, two-photon-subtracted two-mode squeezed vacuum state is more entangled than the single-photon-added two-mode squeezed vacuum state, which is in turn more entangled than the two-mode squeezed vacuum state. The ratio $\tilde{\varepsilon}=2^{\varepsilon}/2^{\varepsilon_{\xi}}$ magnifies the difference between the entanglement content of the photon-added or two-photon-subtracted two-mode squeezed vacuum state with respect to the two-mode squeezed vacuum state. Fig.~\ref{logneg} (b) shows a plot of $\tilde{\varepsilon}$ for both the photon-added and two-photon-subtracted two-mode squeezed vacuum state, as a function of the squeezing parameter $r$.

\section{Quantum teleportation using non-Gaussian entanglement}

We now present some results on CV quantum teleportation using the two-photon-subtracted two-mode squeezed vacuum state, and compare the performance with that of the standard protocol, which uses the two-mode squeezed vacuum state. In particular, we are interested in the teleportation of the non-classical, non-Gaussian Schr{\"o}dinger-cat states $|\Phi_{\rm cat}\rangle$ of Eq.~(\ref{catlikex}).

\begin{figure}[h]\centering
\includegraphics[scale=0.75]{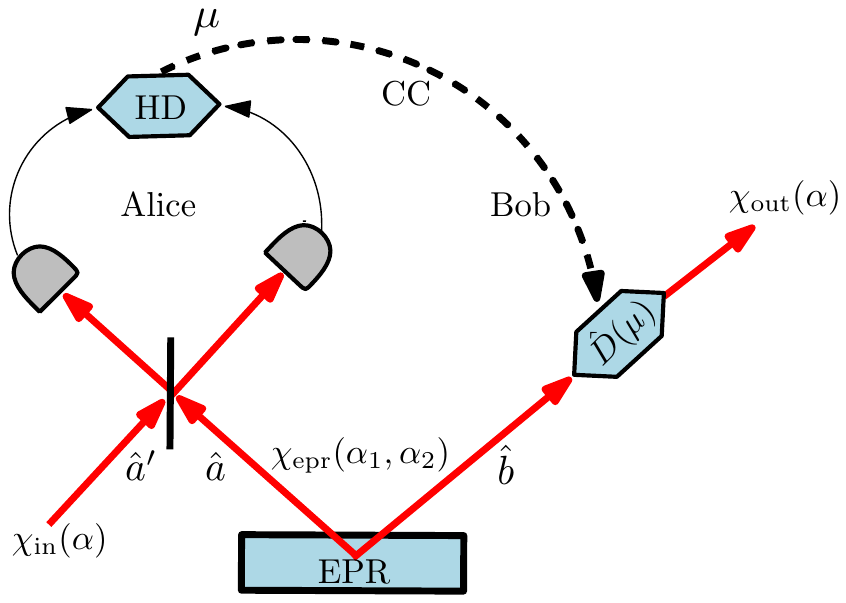}
\caption{(Color online) A schematic of the VBK protocol for CV teleportation. HD, CC stand for homodyne detection and classical communication, respectively. EPR refers to the entangled resource shared between Alice and Bob. In this work, it is either the two-mode squeezed vacuum state or two-photon-subtracted two-mode squeezed vacuum state.}
\label{vbk}
\end{figure}

To begin with, let us briefly describe the standard protocol for CV teleportation, first introduced by VBK (see Fig.~\ref{vbk}). Alice, who wants to transport a single-mode input state to Bob, prearranges the sharing of an entangled resource with him. She mixes the single-mode input state (in mode $\hat{a}'$) with mode $\hat{a}$ of the entangled resource on a 50:50 beam splitter. She then performs a homodyne measurement on the beam-splitter-output modes and classically communicates the result $\mu=q+i p$ to Bob. Assuming balanced homodyning, the real and imaginary parts of $\mu$ satisfy
\begin{equation}
\label{homodyne1}
\frac{1}{\sqrt{2}}\left(\hat{x}_{a'}-\hat{x}_a\right)|q\rangle=q|q\rangle,\  \frac{1}{\sqrt{2}}\left(\hat{y}_{a'}+\hat{y}_a\right)|p\rangle=p|p\rangle,
\end{equation}
where $\hat{x}_a$, $\hat{y}_a$, $\hat{x}_{a'}$ and $\hat{y}_{a'}$ are the canonical operators probed by the homodyne measurement, which are related to the beam-splitter-input mode operators $\hat{a}$, $\hat{a}'$ in a manner similar to the relation given in Eq.~(\ref{homodyne2}). As a final step, Bob performs a displacement operation $\hat{D}(\mu)={\rm exp}(\mu \hat{b}^{\dagger}-\mu^* \hat{b})$ on the mode $\hat{b}$ of the entangled resource, which results in the recovery of the teleported state. 

Mathematically, the above protocol can be described and analyzed in one of many alternative ways~\cite{Braunstein_rmp05}. We adopt an approach based on the use of characteristic functions. Marian and Marian~\cite{Marian_06} showed that, assuming ideal measurements and the case that Alice performs balanced homodyne detection, the Weyl-ordered characteristic function of the teleportation output ($\chi_{\rm out}$) can be written in terms of those of the input ($\chi_{\rm in}$) and the entangled resource ($\chi_{\rm EPR}$) as 
\begin{equation}
\label{charout}
\chi_{\rm out}(\alpha)=\chi_{\rm in}(\alpha)\chi_{\rm EPR}(\alpha^*,\alpha).
\end{equation}
The Weyl-ordered characteristic function is related to the Wigner function via a Fourier transform, as
\begin{equation}
\label{wignercharrel}
W(\beta)=\frac{1}{\pi^2}\int d^2\alpha\chi(\alpha)e^{\beta\alpha^*-\beta^*\alpha}.
\end{equation}
In the case of a single-mode pure state $|\psi\rangle$, it can be easily determined as $\chi(\alpha)=\langle\psi| \hat{D}(\alpha)|\psi\rangle$, where $\hat{D}(\alpha)$ is the displacement operator. The Weyl-ordered characteristic functions of the entangled resources, namely the two-mode squeezed vacuum state and two-photon-subtracted two-mode squeezed vacuum state are given by
\begin{eqnarray}
\label{eprchar}
&\chi_{\xi}(\alpha_1,\alpha_2)=e^{-1/2(|\xi_1|^2+|\xi_2|^2)},&\nonumber\\
&\chi_{{\rm TPS}}(\alpha_1,\alpha_2)= \chi_{\xi}(\alpha_1,\alpha_2)&\nonumber\\
&\times\frac{1-2{\rm Re}\left[e^{-i\phi}\xi_1\xi_2\right]\tanh r +(1-|\xi_1|^2)(1-|\xi_2|^2)\tanh^2 r}{1+\tanh ^2 r},&
\end{eqnarray}
respectively, where
\begin{eqnarray}
\label{expchar}
&\xi_{k}=\alpha_k \cosh r +\alpha_l^*e^{i\phi}\sinh r,\ (k, l=1,2;\ k \neq l).&
\end{eqnarray}

One figure of merit that is commonly used to gauge the performance of teleportation is the fidelity of teleportation $|\langle\psi_{\rm in}|\psi_{\rm out}\rangle|^2$. It can be written in terms of characteristic functions as
\begin{eqnarray}
\label{fidchar1}
\mathcal{F}&=&\frac{1}{\pi}\int{d^2\alpha\chi_{\rm in}(\alpha)\chi_{\rm out}(-\alpha)},\nonumber\\
&=&\frac{1}{\pi}\int{d^2\alpha\chi_{\rm in}(\alpha)\chi_{\rm in}(-\alpha)\chi_{\rm EPR}(-\alpha^*,-\alpha)}.
\end{eqnarray}
For quantum teleportation to preserve some quantum character of the input state, $|\chi_{\rm EPR}|^2$ must have a magnitude bigger than $\exp(-|\alpha|^2/2)$~\cite{DH14, dhnote}. In the case of a non-Gaussian input state, another important figure of merit is the negativity of the Wigner function at the output in comparison to that at the input.

\subsection{Teleportation of Gaussian states}

\begin{table*}[ht]
\centering
\begin{tabular}{|c|c|c|c|}
	\hline
{\bf State} & $\mathcal{F}_1$ & $\mathcal{F}_2$ \\
\hline
$|\alpha_0\rangle$ & $\frac{1}{1+\gamma}$ & $\frac{1+2\gamma+5\gamma^2}{(1+\gamma)^3(1+\gamma^2)}$\\
\hline
$|\xi_0\rangle$ & $\frac{1}{\sqrt{1+2\gamma\cosh2\rho+\gamma^2}}$ & $\frac{1}{4(1+\gamma)^2}\frac{1}{\left(\sqrt{1+2\gamma\cosh2\rho+\gamma^2}\right)^{5}}\times\left[c_4\gamma^4+c_3\gamma^3+c_2\gamma^2+c_1+c_0\right]$\\
	\hline
\end{tabular}
\caption{Teleportation fidelities for a coherent state $|\alpha_0\rangle$ and a single-mode squeezed vacuum state $|\xi_0\rangle$, teleported using the two-mode squeezed vacuum state ($\mathcal{F}_1$) and two-photon-subtracted two-mode squeezed vacuum state ($\mathcal{F}_2$), where $\gamma=e^{-2r}$, $c_4=3\cosh4\rho+8\cosh2\rho+9$, $c_3=2\cosh4\rho+32\cosh2\rho+14$, $c_2=11\cosh4\rho+8\cosh2\rho+21$, $c_1=16\cosh2\rho$ and $c_0=4$.} 
\label{table1} 
\end{table*}

For the sake of completeness, we first present the teleportation of two of the most commonly-used Gaussian states, namely the coherent state and the single-mode squeezed vacuum state~\cite{DAnno_07}. The characteristic functions of the coherent and single-mode squeezed vacuum states are given by
\begin{eqnarray}
\label{cohsqvchar}
&&\chi_{\rm coh}(\alpha;\alpha_0)=\exp\left(\frac{-1}{2}|\alpha|^2+2i{\rm Im}\left[\alpha\alpha_0^*\right]\right),\nonumber\\
&&\chi_{\rm sqv}(\alpha;\rho, \varphi)=\exp\left(\frac{-1}{2}\left|\alpha^*\cosh\rho+e^{-i\varphi}\alpha\sinh\rho\right|^2\right),\nonumber\\
\end{eqnarray}
respectively. The optimal teleportation fidelities for the above states when teleported using the two-mode squeezed vacuum state and two-photon-subtracted two-mode squeezed vacuum state, calculated based on Eq.~(\ref{fidchar1}), are tabulated in Table~\ref{table1}.

\begin{figure}[h]\centering
\includegraphics[scale=0.45]{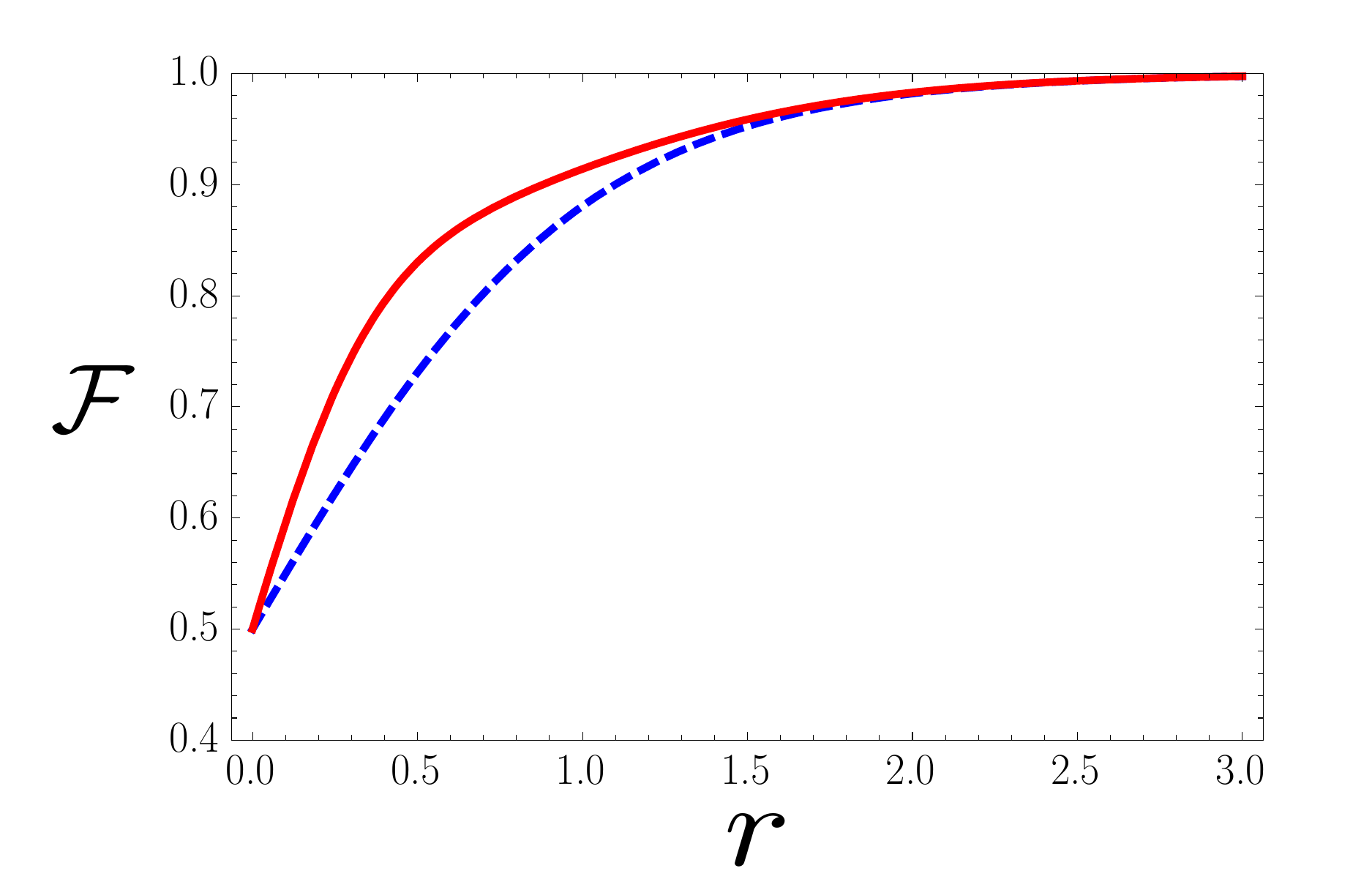}
\caption{(Color online) A plot of the teleportation fidelity $\mathcal{F}$ for coherent states, as a function of the squeezing parameter $r$ of the entangled resources, namely, the two-mode squeezed vacuum state (dashed, blue), two-photon-subtracted two-mode squeezed vacuum state (bold, red).}
\label{cohstatefid}
\end{figure}

\begin{figure}[h]\centering
\includegraphics[scale=0.4]{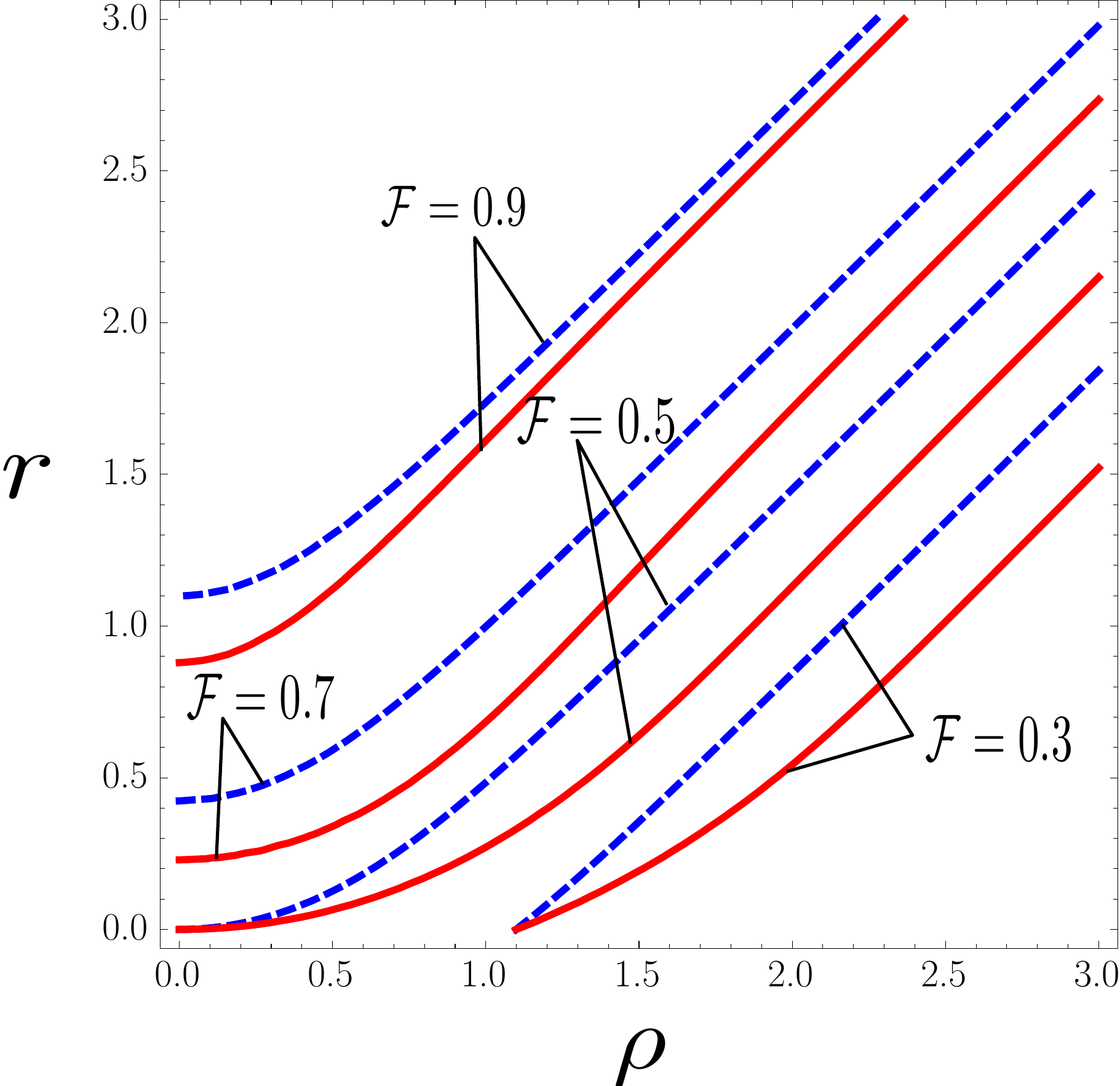}
\caption{(Color online) A contour plot of the teleportation fidelity $\mathcal{F}$ for single mode squeezed vacuum states $|\xi_0\rangle$ of different values of parameter $\rho=|\xi_0|$, for different values of squeezing parameter $r$ of the entangled resources, namely, the two-mode squeezed vacuum state (dashed, blue), two-photon-subtracted two-mode squeezed vacuum state (bold, red).}
\label{sqvfidcon}
\end{figure}

For coherent states, the enhancement to the fidelity of teleportation due to two-photon-subtracted two-mode squeezed vacuum state is independent of the amplitude of the state, as can be seen in Fig.~\ref{cohstatefid}. Figure~\ref{sqvfidcon} presents a contour plot for the fidelity of teleportation of the single mode squeezed vacuum state as a function of the squeezing parameters $r$ and $\rho$. The contours illustrate the fact that at any value of the squeezing parameter $\rho$ of the single-mode squeezed state, two-photon-subtracted two-mode squeezed vacuum state achieves the same fidelity of teleportation as the two-mode squeezed vacuum state while requiring a smaller amount of the squeezing resource. Thus, two-photon-subtracted two-mode squeezed vacuum state offers enhancement over the two-mode squeezed vacuum state in the teleportation of Gaussian states.


\subsection{Teleportation of Schr{\"o}dinger-cat states}

We now discuss the teleportation of non-Gaussian, non-classical Schr{\"o}dinger-cat states. The characteristic function of the cat state is given by
\begin{eqnarray}
\label{catchar}
&\chi_{\Phi}(\alpha;\rho,\varphi)=\left(1-|\tilde{\alpha}|^2\right) e^{-|\tilde{\alpha}|^2/2},&\nonumber\\
 &\tilde{\alpha}=\alpha\cosh\rho-\exp(i\varphi) \alpha^*\sinh\rho, &
\end{eqnarray}
 
 \begin{figure}\centering
\includegraphics[scale=0.45]{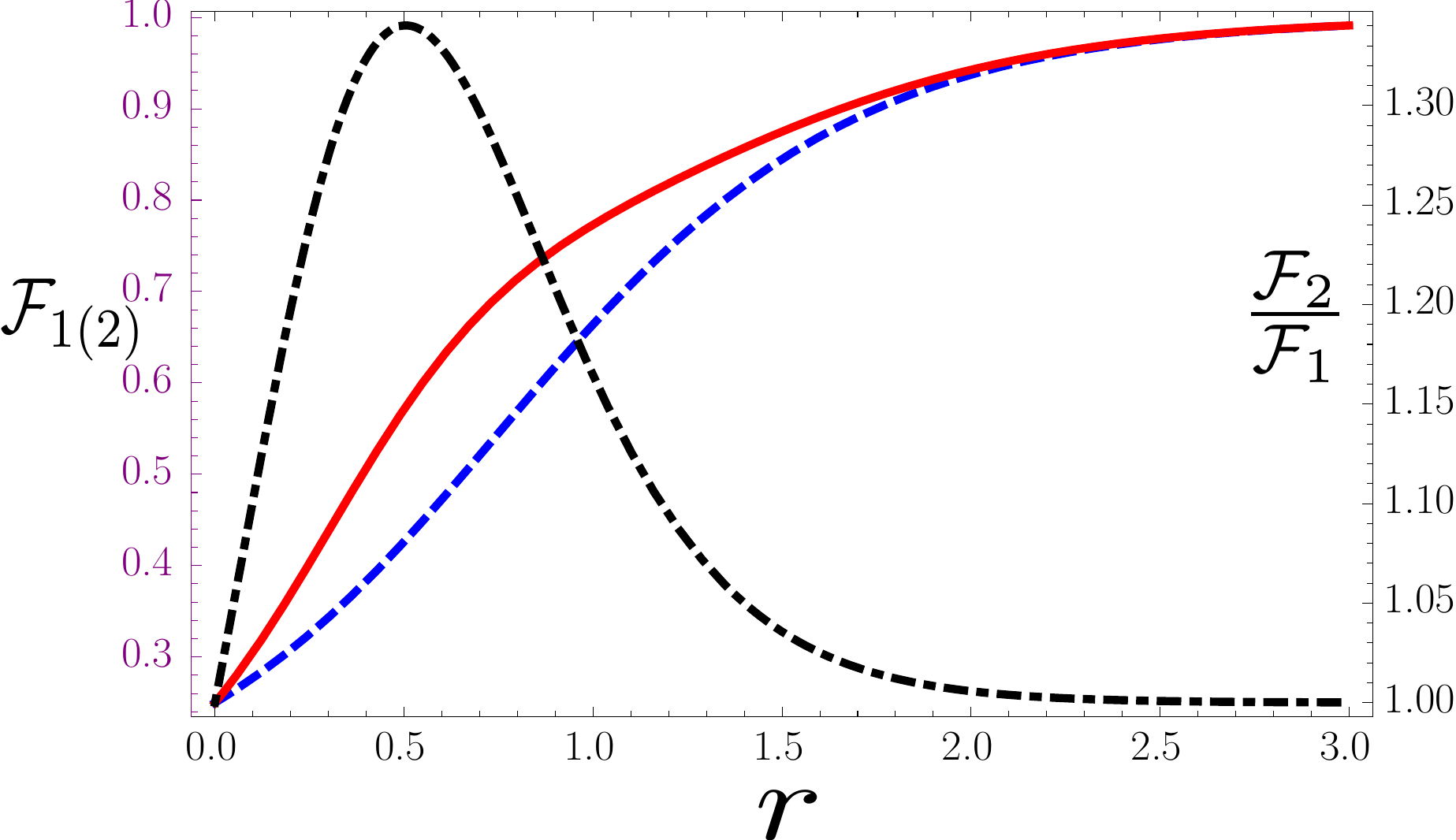}
\caption{(Color online) (Left hand side vertical scale) The teleportation fidelity $\mathcal{F}_{1(2)}$ for the optimal cat state $|\Phi_{\rm cat}\rangle$ corresponding to the state $|\Psi_{\rm cat}\rangle$ with $|\alpha_0|=1$ and $\theta=\pi$, when teleported using the two-mode squeezed vacuum state $\mathcal{F}_{1}$ (dashed, blue), and when teleported using two-photon-subtracted two-mode squeezed vacuum state $\mathcal{F}_{2}$ (solid, red), plotted as a function of the squeezing parameter $r$. (Right hand side vertical scale) The ratio of the two fidelities $\mathcal{F}_2/\mathcal{F}_1$ (dot-dashed, black), plotted as a function of the squeezing parameter $r$.}
\label{fid}
\end{figure}

\begin{figure}[h]\centering
\includegraphics[scale=0.4]{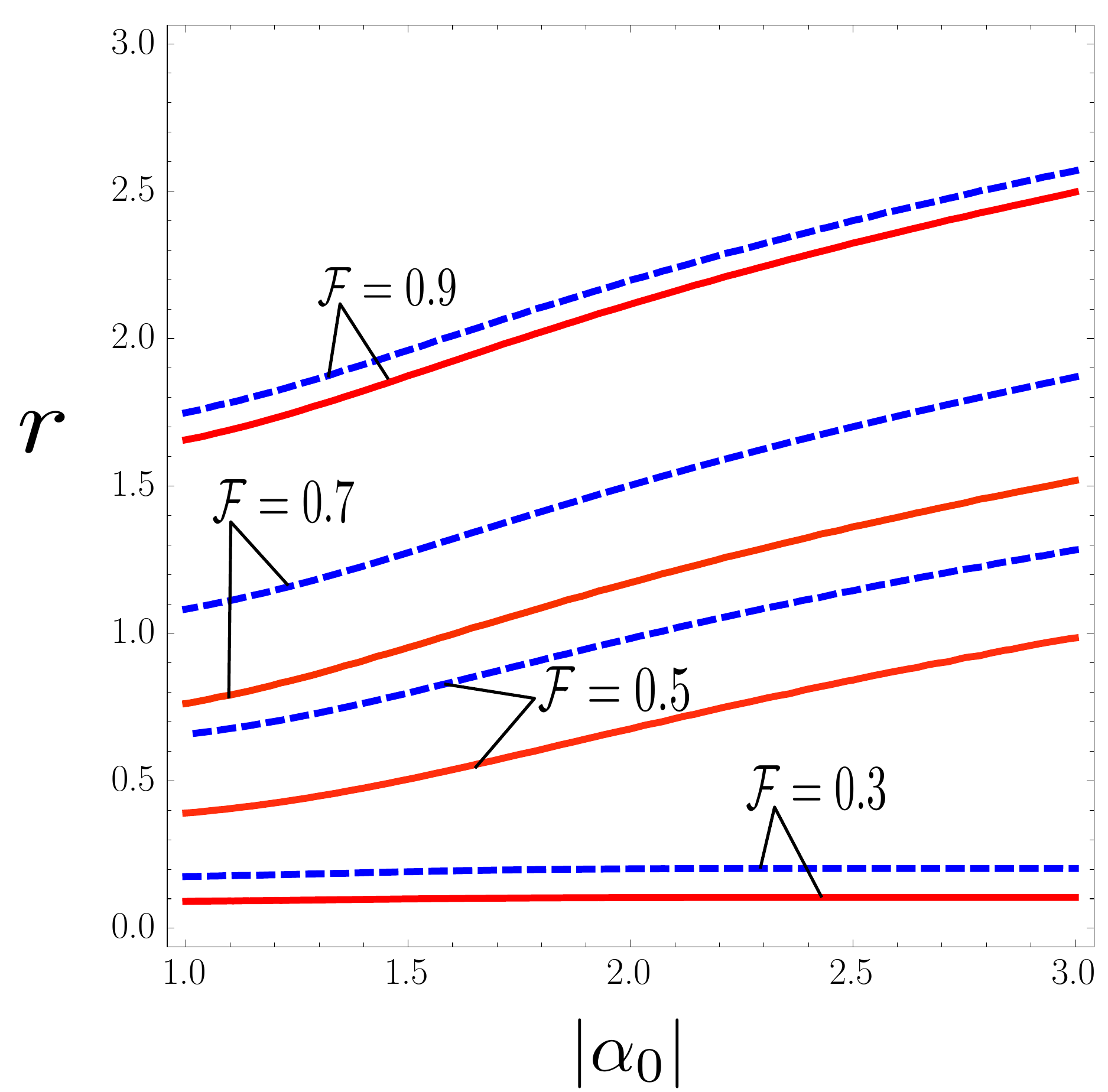}
\caption{(Color online) A contour plot of the teleportation fidelity $\mathcal{F}$ for the optimal cat state $|\Phi_{\rm cat}\rangle$ corresponding to the state $|\Psi_{\rm cat}\rangle$ of different coherent amplitudes $|\alpha_0|$ and $\theta=\pi$, for different values of squeezing parameter $r$ of the entangled resources, namely, the two-mode squeezed vacuum state (dashed, blue), two-photon-subtracted two-mode squeezed vacuum state (solid, red).}
\label{catfidcon}
\end{figure}
 
 \subsubsection{Fidelity of teleportation}
Once again, based on Eq.~(\ref{fidchar1}), the expressions for the teleportation fidelity for the state $|\Phi_{\rm cat}\rangle$, when teleported optimally~\cite{optphi} using the two-mode squeezed vacuum state ($\mathcal{F}_1$) and two-photon-subtracted two-mode squeezed vacuum state ($\mathcal{F}_2$) are calculated, and found to be
\begin{eqnarray}
\label{fid121}
&\mathcal{F}_1(\rho,\varphi,\gamma)=\frac{2+4 \gamma\cosh2\rho+\left(1+3\cosh4\rho\right)\gamma^2+4\gamma^3\cosh2\rho+2\gamma^4}{2 \left(1+2 \gamma  \cosh 2 \rho+\gamma ^2\right)^{5/2}},&\nonumber\\
&\mathcal{F}_2(\rho,\varphi,\gamma)=\hat{\Gamma}\mathcal{F}_1(\rho,\varphi,\gamma),&\nonumber\\ 
&\hat{\Gamma}=\left[1+\frac{\gamma^2(1+\gamma)^2}{2(1+\gamma^2)}\left\{\left(\frac{1-\gamma}{1+\gamma}\right)^2\frac{\partial^2}{\partial\gamma^2}-4\frac{1-\gamma}{(1+\gamma)^2}\frac{\partial}{\partial\gamma}\right\}\right],&
\end{eqnarray}
respectively, where $\gamma=e^{-2r}$. (See Appendix C for the derivation of $\mathcal{F}_2$.)

\begin{figure*}\centering
\includegraphics[scale=0.5]{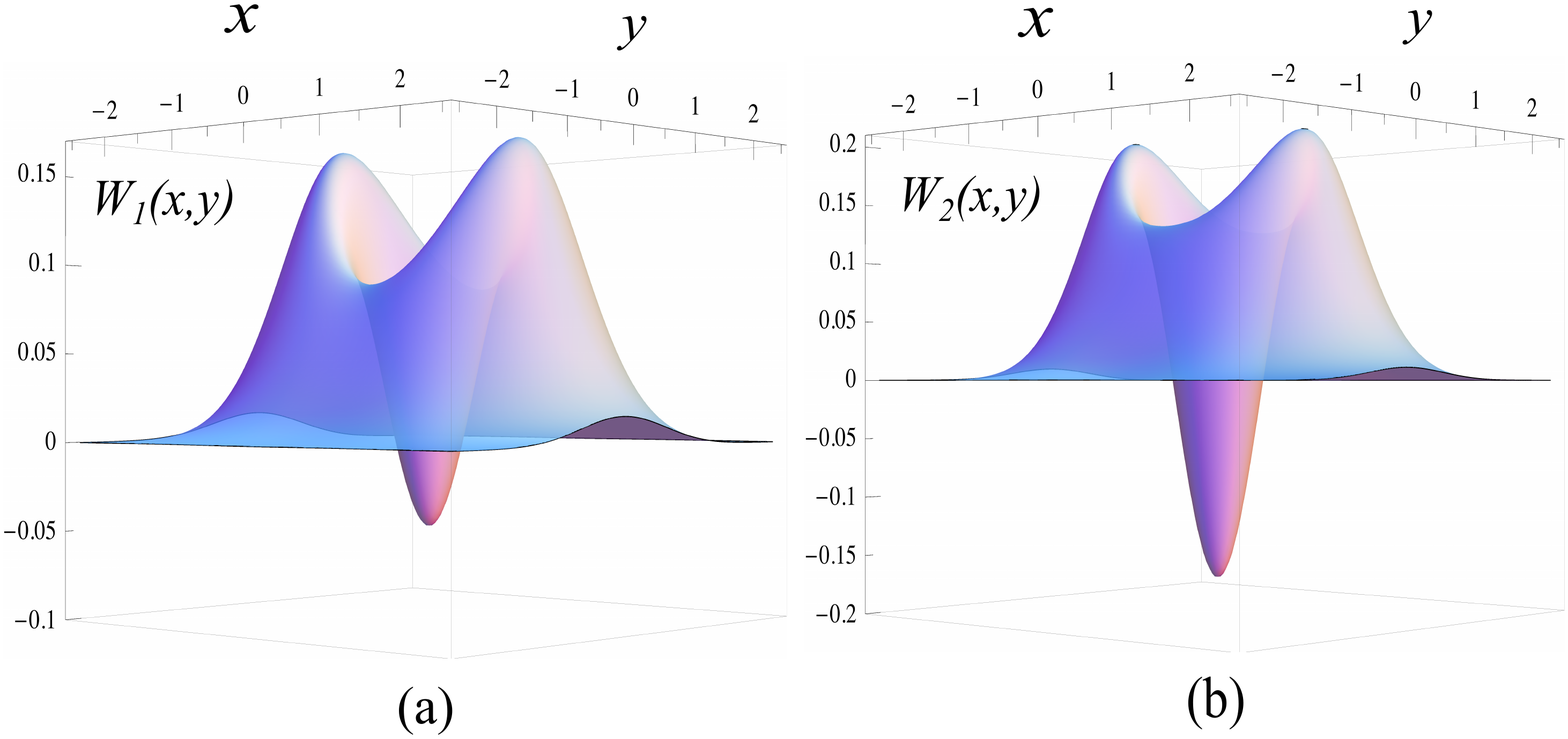}
\caption{(Color online) Wigner function of the optimal cat state $|\Phi_{\rm cat}\rangle$ corresponding to the state $|\Psi_{\rm cat}\rangle$ with $|\alpha_0|=1$ and $\theta=\pi$, at the output of the teleportation process, when teleported using (a) the two-mode squeezed vacuum state (left), (b) two-photon-subtracted two-mode squeezed vacuum state (right).  The squeezing parameter $r$ is chosen to be $0.5$. The Wigner function teleported using two-photon-subtracted two-mode squeezed vacuum state achieves a much lower negative value (0.2) than the value achieved by using the two-mode squeezed vacuum state (0.05).}
\label{wigoddtel}
\end{figure*}

We now focus on the state $|\Phi_{\rm cat}\rangle$ corresponding to $|\Psi_{\rm cat}\rangle$ of coherent amplitude $|\alpha_0|=1$ and $\theta=\pi$. The optimal choice $\rho=0.313$ in Eq.~(\ref{catlikex}) provides an input-state fidelity of $99.7\%$~\cite{Jeong_05, refphase}. Fig.~\ref{fid} shows a plot of the two teleportation fidelities, and their ratio for the teleportation of the above cat state as a function of the squeezing parameter of the entangled resource~$r$. We see that two-photon-subtracted two-mode squeezed vacuum state offers substantially-enhanced fidelity over the two-mode squeezed vacuum state for small values of the squeezing parameter $r$ ($0<r<2$). Figure~\ref{catfidcon} shows a contour plot of the optimal fidelity of teleportation for the cat state $|\Phi_{\rm cat}\rangle$ corresponding to different magnitudes $|\alpha_0|$ of $|\Psi_{\rm cat}\rangle$ with $\theta=\pi$, as a function of the coherent amplitude $|\alpha_0|$ and the squeezing parameter of the entangled resource $r$. (The squeezing parameter $\rho$ for the different input states $|\Phi_{\rm cat}\rangle$ is chosen such that the input-state fidelity is maximized.) The plot elucidates the persistence of the squeezing benefit in using two-photon-subtracted two-mode squeezed vacuum state in place of the two-mode squeezed vacuum state for cat states of Eq.~(\ref{catex}) with increasing values of coherent amplitude $|\alpha_0|$.

\subsubsection{Wigner function negativity}

\begin{figure}[h]\centering
\includegraphics[scale=0.3]{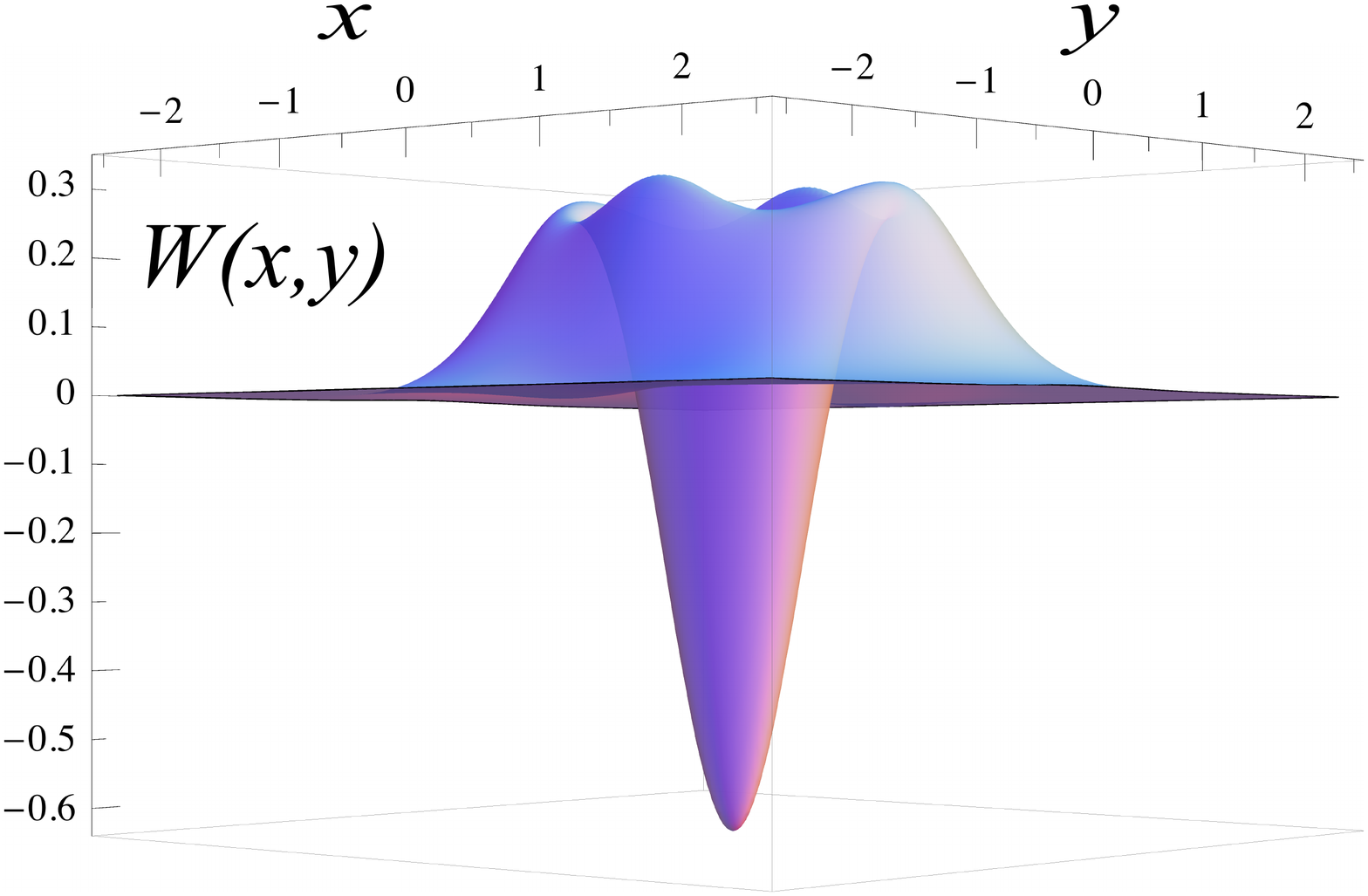}
\caption{(Color online) Wigner function of the optimal cat state $|\Phi_{\rm cat}\rangle$ corresponding to the state $|\Psi_{\rm cat}\rangle$ with $|\alpha_0|=1$ and $\theta=\pi$.}
\label{wigoriginalodd}
\end{figure}

\begin{figure}[h]\centering
\includegraphics[scale=0.45]{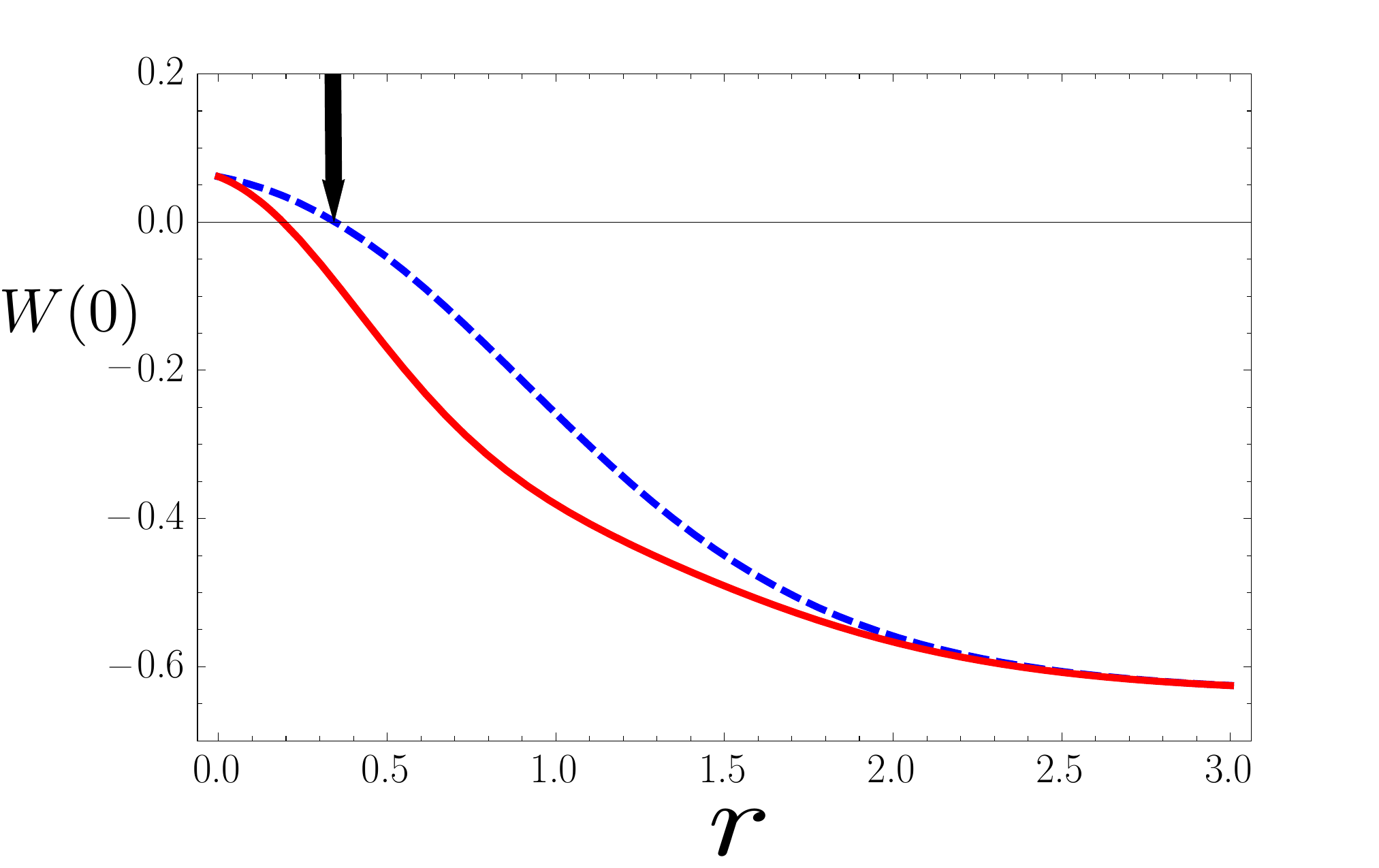}
\caption{(Color online) Wigner function at the phase-space origin for the optimal cat state $|\Phi_{\rm cat}\rangle$ corresponding to the state $|\Psi_{\rm cat}\rangle$ with $|\alpha_0|=1$ and $\theta=\pi$, when teleported using the two-mode squeezed vacuum state (dashed, blue) and two-photon-subtracted two-mode squeezed vacuum state (solid, red). Negative values of the Wigner function of the teleported state are achieved at a smaller value of the squeezing parameter $r$ for two-photon-subtracted two-mode squeezed vacuum state ($r=0.2$, against $r=0.35$ for the two-mode squeezed vacuum state, as indicated by the arrow).}
\label{wigodd0}
\end{figure}

The Wigner function of the state $|\Phi_{\rm cat}\rangle$ of Eq.~(\ref{catlikex}) is given by
\begin{equation}
\label{catwigner}
W(\alpha;\rho,\varphi)=\frac{2}{\pi }  \left(4 \left| \tilde{\alpha} \right|^2-1\right)e^{-2 \left| \tilde{\alpha} \right| ^2},
\end{equation}
where $\tilde{\alpha}=\alpha\cosh\rho-\exp(i\varphi) \alpha^*\sinh\rho$. Based on Eq.~(\ref{wignercharrel}), the Wigner functions at the output of the teleportation process, when teleported optimally~\cite{optphi} using the two-mode squeezed vacuum state ($W_1$) and two-photon-subtracted two-mode squeezed vacuum state ($W_2$), are found to be
\begin{align}
\label{wcatf}
&W_1(\alpha; \rho, \varphi, \gamma)=\frac{2}{\pi(1+4\gamma\cosh2\rho+4\gamma^2)^{5/2}}\nonumber\\
&\times\exp\left(\frac{-2}{1+4\gamma\cosh2\rho+4\gamma^2}(2\gamma\left|\alpha\right|^2+\left|\tilde{\alpha}\right|^2)\right)\nonumber\\
&\times\Big[(4|\tilde{\alpha}|^2-1)+4\left(4\left|\alpha\right|^2-\cosh2\rho\right)\gamma\nonumber\\
&+16\left(3\left|\tilde{\alpha}\right|^2-2\left|\alpha\right|^2\cosh2\rho\right)\gamma^2+16\gamma^3\cosh2\rho+16\gamma^4\Big],&\nonumber\\
&W_2(\alpha; \rho, \varphi, \gamma)=\hat{\Gamma} W_1(\alpha; \rho, \varphi, \gamma),&
\end{align}
respectively, where $\hat{\Gamma}$ is the differential operator given in Eq.~(\ref{fid121}). Figures \ref{wigoriginalodd} and \ref{wigoddtel}, show plots of the Wigner function of the optimal cat state $|\Phi_{\rm cat}\rangle$ corresponding to the state $|\Psi_{\rm cat}\rangle$ with $|\alpha_0|=1$ and $\theta=\pi$, at the input and output of the teleportation process, respectively. The Wigner functions in Fig.~\ref{wigoddtel}(a) and (b) correspond to the output states of the teleportation process with the two-mode squeezed vacuum state and two-photon-subtracted two-mode squeezed vacuum state as the entangled resources, respectively, at a value of the squeezing parameter (of the entangled resource) $r=0.5$. The chosen value of $r$ corresponds to that point at which the ratio of fidelities $\mathcal{F}_2/\mathcal{F}_1$ is maximum in Fig.~\ref{fid}. We find that the Wigner function teleported using two-photon-subtracted two-mode squeezed vacuum state achieves a lower negative value ($0.2$) than the one teleported using the two-mode squeezed vacuum state ($0.05$).

As is obvious from Fig.~\ref{wigoriginalodd}, the Wigner function of the optimal cat state $|\Phi_{\rm cat}\rangle$ corresponding to the state $|\Psi_{\rm cat}\rangle$ with $|\alpha_0|=1$ reaches its maximum negative value at the phase-space origin. Figure~\ref{wigodd0} shows a plot of $W(0)$ of the Wigner function of this state at the output of the teleportation process, as a function of the squeezing parameter of the entangled resource $r$.  The plot illustrates the fact that the threshold value of the squeezing parameter $r$, above which the value of $W(0)$ at the output of the teleportation process becomes negative, is smaller when two-photon-subtracted two-mode squeezed vacuum state is used ($r=0.2$), as compared to the value when the two-mode squeezed vacuum state is used ($r=0.35$). Also, the former becomes more negative than the latter in the range of $0<r<2$.

In this investigation, we have not considered the deterioration of the fidelity due to finite efficiency of the detectors. However, these can be examined by following the standard procedure, {\it e.g.}, as used by Olivares {\it et al.}~\cite{Olivares_03, Olivares_04}.


\section{Summary}

In summary, we discussed the non-Gaussian entanglement that results from the subtraction of a photon from each mode of a two mode squeezed vacuum state. We highlighted many of the characteristic properties of the two-photon-subtracted two-mode squeezed vacuum state. We described a scheme that heralds the state, and discussed how this state can be used as an entangled resource for quantum teleportation. We showed that the two-photon-subtracted two-mode squeezed vacuum state achieves quantum teleportation of the nonclassical Schr{\"o}dinger cat states with a higher fidelity than what can be achieved with two-mode squeezed vacuum light for the same amount of squeezing. Further, we elucidated that the two-photon-subtracted two-mode squeezed vacuum state also achieves a higher maximum negativity of the teleported Wigner function than the two-mode squeezed vacuum state for any given amount of squeezing, thus, enabling better recovery of the nonclassical properties of the teleported state. Quantum teleportation is just one illustration of the usefulness of non-Gaussian entangled states in continuous-variable quantum information. Other applications of such non-Gaussian entangled states include loophole free tests for Bell inequality violations using homodyne detection~\cite{NC04, GFCW04, GPFC05, DK05} and quantum bit commitment that is robust against Gaussian attacks~\cite{MMLC10}---both being instances, where Gaussian states cannot be used, and quantum optical interferometry~\cite{CG12}.


\section{Acknowledgments}
KPS thanks Oklahoma State University for the hospitality during his visit to Stillwater, which is when this work was initiated; the Graduate School of Louisiana State University for the 2014-2015 Dissertation Year Fellowship. JPD would like to acknowledge the support from the AFOSR and the NSF.

\appendix
\section*{Appendix A}
\label{wigcalc}
Consider the expression for the two-photon-subtracted two-mode squeezed vacuum state given in Eq.~(\ref{notoneform}):
\begin{align}
|\psi\rangle_{\rm TPS}&=N\hat{S}\left(\xi\right)\left(|0\rangle_a|0\rangle_b+\eta |1\rangle_a|\ 1\rangle_b\right), \nonumber\\
N&=\frac{1}{\sqrt{1+\tanh^2 r}},\ \eta=e^{i\phi}\tanh r.
\end{align}
Let $\tilde{\rho}$ be the following density matrix:
\begin{align}
\tilde{\rho}&=N^2(|0\rangle\langle0 |_a|0\rangle\langle0|_b+|\eta|^2|1\rangle\langle 1 |_a|1\rangle\langle 1|_b\nonumber\\
&+(\eta^*|0\rangle\langle1 |_a|0\rangle\langle1|_b+{\rm c.c.}))
\end{align}
Its Wigner function can be constructed piece-wise as follows:
\begin{align}
W_{\tilde{\rho}}(\tilde{\alpha}, \tilde{\beta})&=N^2 (W_{|0\rangle\langle0|}(\tilde{\alpha})W_{|0\rangle\langle0|}(\tilde{\beta})\nonumber\\
&+|\eta|^2W_{|1\rangle\langle1|}(\tilde{\alpha})W_{|1\rangle\langle1|}(\tilde{\beta})\nonumber\\
&+(\eta^*W_{|0\rangle\langle1|}(\tilde{\alpha})W_{|0\rangle\langle1|}(\tilde{\beta})+{\rm c.c.})).
\end{align}
The Wigner functions of the vacuum state $|0\rangle\langle0 |_a$ and the single-photon Fock state $|1\rangle\langle1 |_a$ are given by:
\begin{align}
W_{|0\rangle\langle0|}(\tilde{\alpha})&=\frac{2}{\pi}e^{-2|\tilde{\alpha}|^2}\\
W_{|1\rangle\langle1|}(\tilde{\alpha})&=\frac{2}{\pi}e^{-2|\tilde{\alpha}|^2}\left(4|\tilde{\alpha}|^2-1\right).
\end{align}
The Wigner function corresponding to the off-diagonal term $|0\rangle\langle1 |_a$ can be written as:
\begin{align}
W_{|0\rangle\langle1|}(\tilde{\alpha})&=\frac{1}{\pi^2}\int d^2\alpha_1 \langle 1|D_a(\alpha_1)|0\rangle\exp\left(-(\alpha_1\tilde{\alpha}^*-\alpha_1^*\tilde{\alpha})\right)\nonumber\\
&=\frac{1}{\pi^2}\int d^2\alpha_1\ \alpha_1e^{-\frac{1}{2}|\alpha_1|^2}\exp\left(-(\alpha_1\tilde{\alpha}^*-\alpha_1^*\tilde{\alpha})\right)\nonumber\\
&=-\frac{\partial}{\partial\tilde{\alpha}}\left(\frac{2}{\pi}e^{-2|\tilde{\alpha}|^2}\right)\nonumber\\
&=\frac{2}{\pi}e^{-2|\tilde{\alpha}|^2}\times(2\tilde{\alpha}).
\end{align}
Therefore, by combining the different pieces, we can write the Wigner function of $\tilde{\rho}$ as:
\begin{align}
W_{\tilde{\rho}}(\tilde{\alpha},\tilde{\beta})=\left(\frac{2}{\pi}\right)^2 N^2 e^{-2\left(|\tilde{\alpha}|^2+|\tilde{\beta}|^2\right)}\nonumber\\
\times\left(1+4\tilde{\alpha}\tilde{\beta}\left(\eta+\eta^*\right)+\left(4|\tilde{\alpha}|^2-1\right)\left(4|\tilde{\beta}|^2-1\right)|\eta|^2\right).
\label{wtpsfinal}
\end{align}
As the final step, the Wigner function of $\rho=\hat{S}(\xi)\tilde{\rho}\hat{S}^\dagger(\xi)$ can be obtained from Eq.~(\ref{wtpsfinal}) using the change of variables given in Eq.~(\ref{TRANS}).

\section*{Appendix B}
\label{appxa}
\setcounter{section}{1}
The logarithmic negativity $\varepsilon$ of a state can be calculated using the absolute sum of the negative eigenvalues $\mathcal{N}=|\sum_{i}\lambda_i|$, $\lambda_i < 0$ of the partial transpose of its density operator $\rho^{\rm PT}$, as $\varepsilon=\log{\left(1+\mathcal{N}\right)}$. The partial transpose $\rho^{\rm PT}$ corresponding to the state of Eq.~(\ref{ephtmsv1}) is given by
\begin{eqnarray}
\label{ptrans}
\rho^{PT}&=&\sum_{n,m=0}^{\infty}c_n c_m e^{i (n-m) \phi}|n,m\rangle\langle m,n|,\nonumber\\
c_n&=&\frac{\tanh^n r}{\cosh^{3} r\sqrt{1+\tanh^2 r}} (n+1).
\end{eqnarray}
Clearly, the diagonal terms in Eq.~(\ref{ptrans}) are all positive. The off-diagonal terms ($n\neq m$) in Eq.~(\ref{ptrans}) have the form
\begin{eqnarray}
\label{offd}
 |n,m\rangle\langle m,n| e^{i (n-m) \phi}+|m,n\rangle\langle n,m| e^{i (m-n) \phi},
\end{eqnarray}
which can be diagonalized as
\begin{eqnarray}
\label{doffd}
\frac{e^{i n\phi}|n,m\rangle+e^{i m\phi}|m,n\rangle}{\sqrt{2}}\times\frac{e^{-i n\phi}|n,m\rangle+e^{-i m\phi}|m,n\rangle}{\sqrt{2}}\nonumber\\
-\frac{e^{i n\phi}|n,m\rangle-e^{i m\phi}|m,n\rangle}{\sqrt{2}}\times\frac{e^{-i n\phi}|n,m\rangle-e^{-i m\phi}|m,n\rangle}{\sqrt{2}}.\nonumber\\
\end{eqnarray}
Thus, all the negative eigenvalues are $-c_n c_m$, and hence the logarithmic negativity parameter becomes
\begin{eqnarray}
\label{lognegpar}
\varepsilon&=&\log_2\left(1+\sum_{n\neq m}c_n c_m\right),\nonumber\\
&=&\log_2\left(\sum_{n}c_n^2+\sum_{n\neq m}c_n c_m\right),\nonumber\\
&=&\log_2\left(\sum_{n}c_n\right)^2.
\end{eqnarray} 
For the $c_n$ of Eq.~(\ref{ptrans}) corresponding to the two-photon-subtracted two-mode squeezed vacuum state, $\left(\sum_{n}c_n\right)^2=e^{4r}/\cosh2r$, while for the two-mode squeezed vacuum state whose $c_n=\tanh^n r/\cosh r$, $\left(\sum_{n}c_n\right)^2=e^{2r}$.

\section*{Appendix C}
\label{appxb}
\setcounter{section}{1}
The fidelity of teleporation $\mathcal{F}$, as given in Eq.~(\ref{fidchar1}), involves the characteristic function of the entangled resource $\chi_{\rm EPR}(-\alpha^*,-\alpha)$. $\chi_{\xi}(-\alpha^*,-\alpha)$ and $\chi_{{\rm TPS}}(-\alpha^*,-\alpha)$ of Eq.~(\ref{eprchar}), at the optimal value of phase $\phi=\pi$ in Eq.~(\ref{expchar}), take the simplified forms
\begin{eqnarray}
\label{app1}
\chi_{\xi}(-\alpha^*,-\alpha)&=&e^{-\gamma|\alpha|^2},\nonumber\\
\chi_{{\rm TPS}}(-\alpha^*,-\alpha)&=&\frac{1+2\tanh r |\alpha|^2+\tanh^2 r\left(1-|\alpha|^2\right)^2}{1+\tanh ^2 r}\nonumber\\
&\times& e^{-\gamma|\alpha|^2},
\end{eqnarray}
respectively, where $\gamma=e^{-2r}$. The above expression for $\chi_{{\rm TPS}}$ can be written as a differential of $\chi_{\xi}$ with respect to $\gamma$ as:
\begin{equation}
\label{app2}
\chi_{{\rm TPS}}(-\alpha^*,-\alpha)=\hat{\Gamma}\chi_{\xi}(-\alpha^*,-\alpha),
\end{equation}
where $\hat{\Gamma}$ is the differential operator given in Eq.~(\ref{fid121}). In calculating the fidelity of teleportation using two-photon-subtracted two-mode squeezed vacuum state based on Eq.~(\ref{fidchar1}) (and also the Wigner function based on (Eq.~\ref{wignercharrel})), since the integration is independent of the parameter $\gamma$, $\hat{\Gamma}$ can be pulled out of the integral. Hence, the final expression for the fidelity with two-photon-subtracted two-mode squeezed vacuum state differs from that with the two-mode squeezed vacuum state only in having the differential operator $\hat{\Gamma}$ in front.

\bibliographystyle{apsrev}
\bibliography{ref_withouturl}

\begin{thebibliography}{52}
\bibitem[{fid()}]{fidnote}
\bibinfo{note}{Fidelity of teleportation is the overlap of the input state with
  the output state.}
\bibitem[{iof()}]{iofidnote}
\bibinfo{note}{Here, by input and output fidelities, we mean the overlap of the
  input and output states with the cat state $|\Psi_{\rm cat}\rangle$ of
  Eq.~(\ref{catex}), respectively.}
\bibitem[{spa()}]{spatmsv}
\bibinfo{note}{The single-photon-added TMSV can be found in Eqs. (8) and (9) of
  Ref.~\cite{Agarwal_11}.}
\bibitem[{opt()}]{optphi}
\bibinfo{note}{Optimality is achieved when in Eq.~(\ref{expchar}) the squeezing
  angle $\phi$ of the entangled resource is set to $\pi$.}
\bibitem[{ref()}]{refphase}
\bibinfo{note}{The phase $\varphi$ is set to be $0$, in order to fix a phase
  reference.}
\end{thebibliography}

\end{document}